\documentclass[11pt]{article}

\usepackage{color}

\usepackage{amsmath}
\usepackage{amsfonts}
\usepackage{amssymb}
\usepackage{graphicx}
\usepackage[margin=1.0in]{geometry}

\newtheorem{theorem}{Theorem}

\newcommand{\argmax}{\operatorname*{arg\ max}\limits}

\newcommand{\rd}{\ensuremath{\mathrm{d}}}
\newcommand{\id}{\ensuremath{\,\rd}}

\newcommand{\Rset}{\mathbb{R}}
\newcommand{\Id}{\mathcal{I}}
\newcommand{\bigO}{\mathcal{O}}

\newcommand{\MIop}{\textit{I}}
\newcommand{\Hop}{\mathcal{H}}

\newcommand{\barY}{\bar{Y}}

\newcommand{\bx}{{\boldsymbol x}}

\newcommand{\by}{{\boldsymbol y}}
\newcommand{\bc}{{\boldsymbol c}}

\newcommand{\bX}{{\boldsymbol X}}
\newcommand{\sX}{{\mathcal{X}}}
\newcommand{\bh}{{\boldsymbol h}}

\newcommand{\sbb}{{\boldsymbol\beta}}

\newcommand{\sbxi}{{\boldsymbol \xi}}
\newcommand{\sbl}{{\boldsymbol\lambda}}

\newcommand{\bb}{{\boldsymbol b}}
\newcommand{\bk}{{\boldsymbol k}}

\newcommand{\bB}{{\boldsymbol B}}
\newcommand{\bG}{{\boldsymbol G}}
\newcommand{\bH}{{\boldsymbol H}}
\newcommand{\bK}{{\boldsymbol K}}

\title{Sequential Design with Mutual Information for Computer Experiments (MICE):  Emulation of a Tsunami Model}

\author{Joakim Beck\thanks{Department of Statistical Science, University College London, London, UK. ({\em joakim.beck@ucl.ac.uk})} \and Serge Guillas\thanks{Department of Statistical Science, University College London, London, UK. ({\em s.guillas@ucl.ac.uk})}}

\date{}
\begin{document}
\raggedbottom
\maketitle

\begin{abstract}
Computer simulators can be computationally intensive to run over a large number of input values, as required for optimization and various uncertainty quantification tasks. The standard paradigm for the design and analysis of computer experiments is to employ Gaussian random fields to model computer simulators. Gaussian process models are trained on input-output data obtained from simulation runs at various input values. Following this approach, we propose a sequential design algorithm, MICE (Mutual Information for Computer Experiments), that adaptively selects the input values at which to run the computer simulator, in order to maximize the expected information gain (mutual information) over the input space. The superior computational efficiency of the MICE algorithm compared to other algorithms is demonstrated by test functions, and a tsunami simulator with overall gains of up to 20\% in that case.
\end{abstract}

\section{Introduction}
Computer experiments are widely employed to study physical processes \cite{SWN2003,SLC2001}, and involve running a computer simulator which mimics the physical process at various input values. When the computer simulator is computationally expensive to run, say, minutes, hours, or even days, often on a high performance cluster, only a limited number of simulation runs can be afforded, making the planning of such experiments even more important. Surrogate models, also known as emulators, are often used as means for designing and analyzing computer experiments \cite{SWN2003}. Emulators are statistical models that have been used to approximate the input-output behavior of computer simulators for making probabilistic predictions. In this setting, we want to find a design of computer experiments that with minimal computational effort leads to a surrogate model with a good overall fit. We restrict our attention to deterministic computer simulators with a scalar output. In design of experiments it is customary to use space-filling designs \cite{SLC2001}, such as uniform designs, multi-layer designs, maximin(Mm)- and minimax(mM)-distance designs, and Latin hypercube designs (LHD). Space-filling designs treat all regions of the design space as equally important, but are ``one shot'' designs that may waste computations over some unnecessary regions of the input space. A variety of adaptive designs have been proposed which can take advantage of information collected during the experimental design process \cite{LN2008,SWN2003}, typically in the form of input-output data from simulation runs. Some classical adaptive design criteria are the Maximum Mean Squared Prediction Error (MMSPE), the Integrated MSPE (IMSPE), and the entropy criterion (see, e.g., \cite{SWMW1989}). 

We adopt the Design and Analysis of Computer Experiments (DACE) framework proposed in the seminal paper of Sacks et al. \cite{SWMW1989}, within which the computer simulator output is modeled as a realization of a random field, typically assumed Gaussian. When given a set of input-output data, the Best Unbiased Linear Predictor (BLUP), and the associated mean squared prediction error (MSPE), for the random field can be expressed in closed forms \cite{SWMW1989,SWN2003}. Moreover, when the random field is Gaussian, the resulting BLUP is a so-called Gaussian process (GP) emulator. GP emulators are routinely applied to handle computationally intensive computer simulators in the fields of simulation \cite{SWN2003}, global optimization \cite{J2001}, and uncertainty quantification \cite{BZ2012,SGD2012}, among others. Applications include CFD simulation of a rocket booster \cite{GL2009}, and climate simulation \cite{CS2011}. By using the GP approach a range of statistical design criteria can be estimated directly, see \cite{CS2011,SWGO2000}. Finding an optimal design is usually computationally very intensive, except for relatively small designs. A way to circumvent the issue is to consider sequential designs \cite{CS2011,GL2009,LN2008}. In a sequential design, points are systematically chosen, often one at a time. Sequential designs are generally not optimal, but often very effective in practice. An algorithm is called adaptive if it updates its behavior to new data. Two popular sequential designs are Active Learning MacKay (ALM), and Active Learning Cohn (ALC). ALC tends to have better overall predictive performance but involves a higher computational cost \cite{GL2009}.

In this work we propose a new sequential algorithm, called MICE (Mutual Information for Computer Experiments), which is based on the information theoretic mutual information criterion, where the objective of maximizing the information a design provides about the other input values, as suggested by Caselton and Zidek \cite{CZ1984}. Mutual information is a measure of the information contained in one random variable about another \cite{CT2006}. Krause et al. \cite{KSG2008} later proposed a sequential mutual information (MI) algorithm for sensor placement, which sequentially maximizes the mutual information between a GP over the chosen sensor locations and another GP over the locations which have not yet been selected. The MICE criterion is a modified version of the MI criterion in \cite{KSG2008}, where an extra parameter is introduced to improve robustness. This modification is critical when high dimensional spaces are considered. We demonstrate by numerical examples that MICE balances well prediction accuracy and computational complexity. We are particularly interested in deterministic computer simulation experiments with more than just a few input variables.

The article is organized as follows. Section \ref{sec:emulator} reviews Gaussian process modeling for prediction, and present some popular sequential design algorithms within the DACE framework. In Section \ref{sec:3}, a mutual information based design criterion is proposed for computer experiments. The MI algorithm is described in Section \ref{sec:mialg}, and a practical limitation is shown in Section \ref{sec:issue}. Section \ref{sec:micealg} presents the MICE algorithm and some theoretical results. Section \ref{sec:4} details the computational costs associated with the different sequential design algorithms. A numerical comparison of MICE with other methods is provided for a few standard test functions, in lieu of computer simulators, in Section \ref{sec:5}, and for a tsunami simulator that solves nonlinear shallow water equations in Section \ref{sec:6}. Critically, we examine accuracy versus computational cost, as some algorithms can be quite time consuming. Section \ref{sec:7} summarizes our conclusions. Proofs of theorems are in the appendix.

\section{Gaussian process modeling for prediction}
\label{sec:emulator}
We here follow the approach proposed by Sacks et al. \cite{SWMW1989}, where a deterministic computer simulator $y(\bx): \sX \subseteq \Rset^p \to \Rset$ is treated as a random function, $Y(\bx)$, $\bx \in \sX$, except at the points where the simulator output is known. More specifically, $Y(\bx)$ is modeled as a random field with $E\left( Y^2(\bx) \right) < \infty$ given a set of training data, which consists of $n$ input-output pairs $(\bX,\by)$, where $\bX=\left( \bx_j \right)^n_{j=1}$, $\by=\left(y_j\right)^n_{j=1}$, and $y_j=y(\bx_j)$. 

The aim is to determine a random process that can describe the set of data sufficiently well. It is customary that the mean $\operatorname{E}[Y(\bx)]$ takes the form $\bh^T(\bx)\sbb$, that is, a linear combination of $q$ regressors $\bh(\bx): \sX \to \Rset^{q}$ with coefficients $\sbb \in \Rset^q$. In practice, a fixed constant, or a linear regression model, tends to perform well. The covariance $\operatorname{Cov}(Y(\bx),Y(\bx'))$, for $\bx,\bx' \in \sX$, is written in the form $\Sigma(\bx,\bx';\sigma^2)=\sigma^2K(\bx,\bx')$, where $\sigma^2(>0)$ is a scale parameter (often called the process variance) and $K(\bx,\bx')$ is the correlation function. The correlation function is often expressed as a product of stationary, one-dimensional correlation functions. One such choice is the squared-exponential (SE) correlation \cite{RW2006}: 
\begin{equation}
K(\bx,\bx';\sbxi)=\prod_{i=1}^p \exp\left(\frac{-(x_i-x_i')^2}{2\ell_{i}^2} \right), 
\end{equation}
where $\sbxi=(\ell_1,\ell_2,\ldots,\ell_p)^T \in \Rset^p_{+}$. Here $\ell_{i}$ represents the correlation length for the $i$-th input dimension.

In this approach, for predicting the output $y(\bx)$ at any desired $\bx \in \sX$, linear predictors are considered of the form $\hat{y}(\bx)=\sbl^T(\bx)\by$ for some vector $\sbl(\bx) \in \Rset^{n}$. The Best Linear Unbiased Predictor (BLUP), assuming $\sbxi$ is known, is the one that minimizes the Mean Squared Prediction Error (MSPE) with respect to $\sbl(\bx)$,
\begin{equation}
\label{eq:msecond}
\operatorname{MSE}[\sbl^T(\bx)\by]=E\left[(\sbl^T(\bx)\by-Y(\bx))^2\right],
\end{equation}
subject to the unbiasedness constraint $\bH^T\sbl(\bx)=\bh(\bx)$, where $\bH=\left( \bh(\bx_j) \right)_{j=1}^n$. The MSPE of $\hat{y}(\bx)$ is minimized for
\begin{equation}
\hat{\lambda}(\bx)=\bk^T(\bx)\bK^{-1}+\bK^{-1}\bH\dfrac{\bh(\bx)-\bH^T\bK^{-1}\bk(\bx)}{\bH^T\bK^{-1}\bH},
\end{equation}
which leads to the BLUP of $Y(\bx)$:
\begin{align}
\label{eq:predmean}
\hat{y}(\bx) & = \sbl^T(\bx)\by=\bk^T(\bx)\bK^{-1}\by+\bK^{-1}\bH\dfrac{\bh(\bx)-\bH^T\bK^{-1}\bk(\bx)}{\bH^T\bK^{-1}\bH}\by \\
& = \bh^T(\bx)\hat{\sbb}+\bk^T(\bx)\bK^{-1}(\by-\bH\hat{\sbb}), \nonumber
\end{align}
where $\hat{\sbb}=(\bH^T\bK^{-1}\bH)^{-1}\bH^{T}\bK^{-1}\by$ is the generalized least squares estimate of $\sbb$, $\bK$ is the $n \times n$ correlation matrix whose $(i,j)$-th entry is given by $K(\bx_i,\bx_j;\sbxi)$ for $\bx_i,\bx_j \in \bX$, and the $n \times 1$ vector $\bk(\bx;\sbxi)$ has entry $j$ given by $K(\bx,\bx_j;\sbxi)$ for $\bx_j \in \bX$. The correlation matrix must be positive semidefinite. The MSPE is given by:
\begin{align}
\label{eq:predmse}
\operatorname{MSE}[\hat{y}(\bx)] & = \sigma^2(1-\bk^T(\bx)\bK^{-1}\bk(\bx) \\ 
\quad & +\dfrac{(\bh(\bx)-\bH^T\bK^{-1}\bk(\bx))^T(\bh(\bx)-\bH^T\bK^{-1}\bk(\bx))}{\bH^T\bK^{-1}\bH}). \nonumber
\end{align}
The predictor is unbiased, and interpolates the training data, that is, $\hat{y}(\bx_j)=y(\bx_j)$ for $\bx_j \in \bX$. Note that the regularity of the correlation function $K(\bx,\bx)$ determines the regularity of the predictor $\hat{y}(\bx)$ \cite{S1999}, which means that the regularity of $y(\bx)$ should ideally be reflected in the choice of correlation structure. 

As in \cite{SWMW1989}, we also make the assumption that $Y(\bx)$ is a Gaussian process (GP), which is convenient from a computational perspective. This yields a GP emulator of $y(\bx)$ \cite{R1991} with mean $\hat{y}(\bx)$ and variance
\begin{equation}
\label{eq:predvar}
\hat{s}^2(\bx)=\operatorname{MSE}[\hat{y}(\bx)].
\end{equation}
Here $\operatorname{MSE}[\hat{y}(\bx)]$ may be viewed as a measure of uncertainty in the prediction. A GP with the SE correlation function is infinitely mean square differentiable, and the realizations (or sample paths) of this process tend to be unrealistically smooth for modeling computer experiments \cite{S1999}. To be more general, we consider the Mat\'ern family of correlation functions \cite{HS1993}:
\begin{equation}
K(\bx,\bx';\sbxi)=\prod_{i=1}^p \dfrac{1}{2^{\nu-1}\Gamma_\nu}\left( \dfrac{2\nu^{\frac{1}{2}}\vert x_i-x_i' \vert}{\ell_i} \right)^{\nu}J_\nu\left( \dfrac{2\nu^{\frac{1}{2}}\vert x_i-x_i' \vert}{\ell_i} \right),
\end{equation}
where $\sbxi=(\ell_1,\ldots,\ell_p,\nu)^T$, $\Gamma_\nu$ is the Gamma function for $\nu$, and $J_\nu$ a modified Bessel function of order $\nu>0$. The parameter $\nu$ regulates the smoothness of the process, which allows us to model data of different degrees of smoothness. The SE correlation is a special case of a Mat\'ern correlation when $\nu$ goes to $\infty$. A GP with the Mat\'ern correlation function is $\lfloor \nu-1 \rfloor$ times mean square differentiable \cite{S1999}, where $\lfloor \: \rfloor$ denotes the floor function. The Mat\'ern correlation function with fixed $\nu=5/2$, which can be written in an explicit form \cite{RW2006}, is the one used in our numerical tests (unless stated otherwise).

\subsection{Maximum likelihood estimation of unknown parameters}
\label{sec:uncertain}
The parameters involved in the covariance structure are usually unknown ($\sigma^2$ and $\sbxi$, say) and need to be estimated. In this work, the parameters are estimated by maximum likelihood estimation (MLE) using available input-output data (see, e.g., \cite{SWN2003} and references therein). The MLE of $\sigma^2$ is $\hat{\sigma}^2(\sbxi)=(\by-\bH\hat{\sbb}(\sbxi))^T\bK_{\sbxi}^{-1}(\by-\bH\hat{\sbb}(\sbxi))/n$ for fixed $\sbxi$ \cite{SWMW1989}, and the MLE of $\sbxi$, denoted by $\hat{\sbxi}$, can be found by maximizing the profile log-likelihood:
\begin{equation}
\label{eq:mle}
\hat{\sbxi}=\argmax_{\sbxi \in \Xi} \mathcal{L}_p(\sbxi),
\end{equation} 
where $\mathcal{L}_{p}(\sbxi)=\mathcal{L}(\hat{\sigma}^2(\sbxi),\hat{\beta}(\sbxi),\sbxi)$ is the profile log-likelihood for $\sbxi$, $\mathcal{L}(\cdot)$ is the marginal log-likelihood function, and $\Xi$ is a search domain. Assuming the data are normally distributed, the negative log-marginal likelihood is
\begin{equation}
\label{eq:likelihood}
-\mathcal{L} = \frac{1}{2} \ln \lvert \bK \rvert + \frac{1}{2}\bH^T\bK^{-1}\bH+\frac{n}{2}\log{2\pi\sigma^2},
\end{equation}
which means optimization problem \eqref{eq:mle} can be solved by finding the values of $\sbxi$ that maximize $n\ln \hat{\sigma}^2(\sbxi) + \ln \operatorname{det}(\bK_{\sbxi})$, see \cite{SWN2003}. By inserting the MLEs as if they were the \emph{true} values, we have the so-called \emph{estimated} BLUP (EBLUP) $\hat{y}(\bx;\hat{\sbxi})$ \cite{ZC1992}. As shown in \cite{ZC1992}, the estimator $\operatorname{MSE}[\hat{y}(\bx;\hat{\sbxi})]$ tends to underestimate the MSPE.

A Bayesian approach to model the uncertain parameters could have been considered, see, for example, \cite{HS1993,RW2006}, but at a higher computational cost. Note that, although we restrict our attention to MLE, our results are still relevant within a Bayesian setting.

\subsection{The design of computer experiments}
\label{sec:3}
This section presents some of the approaches to the design of computer experiments where the goal is to determine at which input values should data be collected to predict the computer simulator values over the design space $\sX \subseteq \Rset^p$. There are a variety of ways to design such experiments \cite{LN2008,SWN2003}. Design criteria based on the MSPE are natural choices \cite{SWMW1989}. For example, the Maximum MSPE (MMSPE) criterion $\max_{\bx \in \sX} \operatorname{MSE}[\hat{y}_N(\bx)]$, and the Integrated MSPE (IMSPE) criterion $\int_{\sX} \operatorname{MSE}[\hat{y}_N(\bx)] \id \bx$, both to be minimized. Here the subscript $N$ denotes the number of design points in the training data. When $\sX$ is not discrete, the optimization search in $\sX$ is a rather formidable task. In practice, when continuous, $\sX$ is often discretized into a finite grid, $\bX_{G}$, with $N_G$ number of points. Consequently, we replace the search over $\sX$ by a search over a set of candidate points $\bX_{cand} \subseteq \bX_{G}$.

There are also criteria based on information entropy (that is, the negative measure of information \cite{CT2006}). For instance, Lindley \cite{L1956} proposed that the expected change in entropy can serve as a criterion for design. This criterion has been applied by Currin et al. \cite{CMMY1988} to the design of computer experiments.  The \emph{entropy} of a random vector $\barY_N=\barY[\bX_N]=[Y(\bx_1),Y(\bx_2),\ldots,Y(\bx_n)]$ with joint probability distribution $p_{\barY_n}(\by)$ is defined (in bits) as: 
\begin{equation}
\label{eq:shannon}
\Hop(\barY_n)=E[-\log_2(\barY_n)] = -\int\int\cdots\int \log_2(p_{\barY_n}(\by))p_{\barY_n}(\by) \id \by.
\end{equation}
When $Y(\bx)$ is a GP with correlation matrix $\bK$, we obtain the explicit entropy of $\barY_n$:
\begin{equation}
\label{eq:hlog2}
\Hop(\barY_n)=\frac{1}{2}\log_2(2\pi\operatorname{e})^n\operatorname{det}(\bK).
\end{equation}
Maximum entropy sampling \cite{SW1987} uses the entropy criterion to choose the subset of size $N$ of highest entropy, that is,
\begin{equation}
\label{eq:entropyopt}
\argmax_{\bX_N \subset \bX_{cand}} \Hop(\barY_N),
\end{equation}
wherein $\barY_N=\barY[\bX_N]$. Finding the exact solution to optimization problem \eqref{eq:entropyopt} is NP-hard \cite{KLQ1995}. 

We consider sequential designs as practical, computationally cheaper alternatives to ``one shot'' designs, albeit often suboptimal. The \emph{sequential design} is defined as follows: Suppose that we have an initial design $D_n=\{(\bx_j,y_j)\}^n_{j=1}$, then for each $k=n,n+1,n+2,\ldots$ one collects an input-output pair $(\bx_{k+1},y(\bx_{k+1}))$ by choosing the input values
\begin{equation}
\label{eq:seqdesign}
\bx_{k+1}=\argmax_{\bx \in \bX_{cand} \backslash \bX_{k}} f_{k}(\bx),
\end{equation}
wherein $f_{k}(\cdot)$ is a design criterion to be maximized. The algorithm iterates until a stopping criterion is met, or the computational budget allocated is exhausted. The initial design $D_{n}$ could be the empty set $\emptyset$. The sequential design allows sequential acquisition of new design point, and is called \emph{adaptive} if $f_k(\cdot)$ exploits information provided by the collected design $D_{k}=(\bX_k,\by_k)=(\bx_j,y(\bx_j))^{k}_{j=1}$, by, for example, maximum likelihood. The ability to adapt is why sequential designs often outperform one-stage designs such as LHDs. In our context, the covariance parameters need to be estimated. Sequential designs allow the estimates to be improved sequentially with the addition of new design points. This is especially advantageous when some input variables are considerably more influential on the output of interest than others.

Two popular sequential designs for computer experiments are Active Learning MacKay (ALM) and Active Learning Cohn (ALC) \cite{GL2009}. Under the GP assumption, ALM and ALC can be viewed as sequential versions of MMSPE and IMSPE, respectively. There are also other more recent criteria, for instance, Lam and Notz \cite{LN2008} developed the expected improvement for global fit (EIGF) criterion, inspired by a modified expected improvement criterion for global optimization \cite{S1998}. It utilizes the nearest known design point  (in Euclidean distance) to estimate the expected improvement in fit. As shown in \cite{KJ2011}, EIGF can perform better than several well established methods, including ALM, when the output is highly non stationary, as it strongly relies on local information. However, in their study whenever the output behavior were essentially stationarity, EIFG performed worse.

\subsubsection{The ALM algorithm}
At stage $k$ in the sequential design, ALM chooses the design point $\bx_{k+1}$ that maximizes the predictive variance, Eq. \eqref{eq:predvar}, of the GP:
\begin{equation}
\label{eq:alm}
\bx_{k+1}= \argmax_{\bx \in \bX_{cand}} \hat{s}^2_{k}(\bx).
\end{equation}
ALM places many points on the boundary of the design region, especially in the beginning of the selection process. Some argue that boundary points generally are less ``informative'' than nearby interior points, see \cite{KSG2008}. The number of boundary points grows rapidly with the dimension size $p$. Suppose that we have a regular grid with $N^p$ points, then the ratio of boundary points to the total number is $(1-(1-2/N)^p)$. For example, if $p=4$ and $N=10$ the ratio is about $0.59$, and if $p=6$ and $N=10$, nearly $0.74$.

\subsubsection{The ALC algorithm} ALC chooses the design point $\bx_{k+1}$ that yields the largest expected reduction in predictive variance over the design space, and is defined as:
\begin{equation}
\label{eq:alc}
\bx_{k+1}=\argmax_{\bx \in \bX_{cand}} \int_{\sX} \left( \hat{s}^2_{k}(\bx')-\hat{s}^2_{k \cup \bx}(\bx') \right) \id \bx'.
\end{equation}
Standard practice is to approximate the integral over $\sX$ with an average over a grid of $N_{ref}$ reference points in the design space, that is,
\begin{equation}
\label{eq:alcapprox}
\bx_{k+1}=\argmax_{\bx \in \bX_{cand}} \frac{1}{N_{ref}} \sum_{i=1}^{N_{ref}} \left( \hat{s}^2_{k}(\bx_i)-\hat{s}^2_{k \cup \bx}(\bx_i) \right).
\end{equation}

For each $\bx \in \bX_{cand}$, a Cholesky decomposition of $\bK_{k \cup \bx}$ is computed, resulting in a time complexity of $\bigO(N_{cand}N_{ref}k^3)$ for ALC. The computational complexity of step $k$ in ALC can be reduced further from $\bigO(N_{cand}N_{ref}k^3)$ to $\bigO(k^3+N_{cand}N_{ref}k^2)$, by adopting the implementation used in \cite{GL2009} that is based on the following calculations: First, $\bK_{k}^{-1}$ is obtained in $\bigO(k^3)$, and then $\bK_{k \cup \bx}^{-1}$ is computed in $\bigO(k^2)$ by exploiting that $\bK_{k \cup \bx}^{-1}$ can be expressed in terms of $\bK_{k}^{-1}$ and $\bk_{k}(\bx)$:
\begin{equation}
\label{eq:invrel}
\bK_{k\cup \bx}^{-1} = \left( \begin{array}{cc}
\bK_{k}^{-1}+\frac{1}{c}\bK_{k}^{-1}\bk_{k}(\bx)\bk_{k}^T(\bx)\bK_{k}^{-1} & -\frac{1}{c}\bK_{k}^{-1}\bk_{k}(\bx) \\
-\frac{1}{c}\bk_{k}^T(\bx)\bK_{k}^{-1} & \frac{1}{c} \end{array} \right),
\end{equation}
where $c=1-\bk_{k}^T(\bx)\bK_{k}^{-1}\bk_{k}(\bx)$. Next, as shown in \cite{GL2009}, the ALC solution can be obtained by solving the following problem in $\bigO(k^3+N_{cand}N_{ref}k^2)$:
\begin{equation}
\label{eq:alcfinal}
\bx_{k+1}=\argmax_{\bx \in \bX_{cand}} \frac{1}{N_{ref}} \sum_{i=1}^{N_{ref}} \dfrac{V_{k}^2(\bx,\bx_i)}{\hat{s}^2_{k}(\bx)},
\end{equation}
where
\begin{align}
\label{eq:predcovar}
V_{k}(\bx,\bx_i) & =\sigma^2(1-\bk_k^T(\bx)\bK_k^{-1}\bk_k(\bx_i)  \\
\quad & +\dfrac{(\bh(\bx)-\bH^T\bK_k^{-1}\bk_k(\bx))^T(\bh(\bx_i)-\bH^T\bK_k^{-1}\bk_k(\bx_i))}{\bH^T\bK_k^{-1}\bH}). \nonumber
\end{align}

ALC tends to provide a better global fit than ALM for a fixed design size \cite{GL2009,SWGO2000}. ALM is on the other hand easy to implement and relatively cheap computationally, and for this reason often preferred over ALC, see, e.g., \cite{BZ2012}.

\section{Mutual information for the design of computer experiments}
\label{sec:mice}
Mutual information, which, like entropy, is a classical information theoretic measure \cite{CT2006}. It has been used for sensor network design \cite{CZ1984,KSG2008}, experimental design \cite{HM2013}, and optimization \cite{CPV2014}. This section begins with a brief account of mutual information-based design algorithms. Then, in Section \ref{sec:micealg}, we are proposing a new sequential design algorithm based on mutual information.

Suppose that we have two random vectors $\barY$ and $\barY'$ with marginal probability density functions (pdfs) $p_{\barY}(\by)$ and $p_{\barY'}(\by')$, and joint pdf $p_{\barY,\barY'}(\by,\by')$, Then the relationship between \emph{mutual information} of the two vectors, denoted by $\MIop(\barY;\barY')$, and entropy can be written as follows \cite{CT2006}:
\begin{equation}
\label{eq:MIent}
\MIop(\barY;\barY')=\Hop(\barY)-\Hop(\barY \vert \barY').
\end{equation}
The mutual information is equivalent to the Kullback-Leibler divergence between $p_{\barY,\barY'}$ and $p_{\barY}p_{\barY'}$ \cite{CT2006}:
\begin{equation}
\MIop(\barY;\barY')=\int \int \ldots \int \log{\left( \frac{p_{\barY,\barY'}(\by,\by')}{p_{\barY}(\by)p_{\barY'}(\by')} \right)} p_{\barY,\barY'}(\by,\by') \id\by\id\by',
\end{equation}
with $\log(0)0=0$. Caselton and Zidek \cite{CZ1984} showed that mutual information can be utilized to design sampling networks, by choosing the design matrix $\bX_N^* \subset \Rset^{N \times p}$ that maximizes the mutual information between $\barY[\bX_{N}^*]$ and $\barY[\bX_G \backslash \bX_{N}^*]$, that is,
\begin{equation}
\label{eq:czproblem}
\bX_N^*=\argmax_{\bX_N \subset \bX_{cand}} \MIop(\barY[\bX_G \backslash \bX_{N}];\barY[\bX_N]),
\end{equation}
where $\bX_G$ is a discrete design space, and $\bX_{cand} \subseteq \bX_G$ is the set of candidate points available for selection. In other words, the objective is to select the set $\bX_N^*$ that reduces the entropy over $\bX_G \backslash \bX_{N}^*$ the most. This optimization problem is NP-hard \cite{KSG2008}.

\subsection{The MI algorithm}
\label{sec:mialg}
Krause et al. \cite{KSG2008} presented an alternative to avoid the need to directly solve optimization problem \eqref{eq:czproblem}, namely, a sequential algorithm that maximizes the difference $\MIop(\barY[\bX_k \cup \bx];\barY[\bX_G \backslash (\bX_{k} \cup \bx)])-\MIop(\barY[\bX_k];\barY[\bX_G \backslash \bX_{k}])$ with respect to $\bx \in \bX_{cand}$, at each stage $k$ in the sequential design. By adopting the GP approach, as described in Section \ref{sec:emulator}, they have also shown that this optimization problem can be written as:
\begin{equation}
\label{eq:micriterion}
\argmax_{\bx \in \bX_{cand}} \Hop(Y(\bx) \vert \barY_k)-\Hop(Y(\bx) \vert \barY_{G \backslash (k \cup \bx)}) = \argmax_{\bx \in \bX_{cand}} \hat{s}^2_{k}(\bx)/\hat{s}^2_{G \backslash (k \cup \bx)}(\bx),
\end{equation}
since
\begin{align}
\label{eq:micriterionnew}
\Hop(Y(\bx) \vert \barY_k)-\Hop(Y(\bx) \vert \barY_{G \backslash (k \cup \bx)}) & = \frac{1}{2}\log{\left( 2\pi \operatorname{e} \hat{s}^2_{k}(\bx) \right)}-\frac{1}{2}\log{\left( 2\pi \operatorname{e} \hat{s}^2_{G \backslash (k \cup \bx)}(\bx) \right)} \nonumber \\
\quad & \propto \ \hat{s}^2_{k}(\bx)/\hat{s}^2_{G \backslash (k \cup \bx)}(\bx).
\end{align}
Here $G \backslash (k \cup \bx)$ denotes $\bX_G \backslash (\bX_k \cup \bx)$. Note that the objective in optimization problem \eqref{eq:micriterion} has a closed-form expression. This is the greedy mutual information (MI) criterion, or in short the MI criterion. This greedy formulation provides a constant-factor approximation of the original optimization problem \eqref{eq:czproblem} under some mild conditions \cite{KSG2008}. More specifically, the approximation is within $1-1/e$ of the optimum, provided that certain regularity assumptions are satisfied, and the spacing between the points in $\bX_G$ is not too large (see Corollary 6 and Theorem 7 in \cite{KSG2008}). Moreover, the proof exploits that mutual information is a submodular function \cite{NWF1978}; more specifically, that the set function $\MIop(\barY[\bX];\barY[\bX'])$ is submodular for any $\bX,\bX' \subseteq \bX_G$, with $\MIop(\emptyset;\barY[\bX_G])=0$. Greedy algorithms are known to be quite efficient for submodular set functions. The \emph{MI algorithm} proposed by Krause et al. \cite{KSG2008} is given below:\\

\textbf{MI algorithm:} \\
\fbox{\parbox{0.95\textwidth}{
\begin{itemize}
\item[] \textbf{Require:} GP emulator $(\bh(\cdot),K(\cdot,\cdot;\sbxi))$, nugget parameter $\tau^2$, grid $\bX_G$, candidate set $\bX_{cand} \subseteq \bX_G$, a design $\bX_{k} \subset \bX_{cand}$ of size $k$, desired design size $N$
\item[] \textbf{Step 1.} Let $\bX_{cand} \leftarrow \bX_{cand} \backslash \bX_{k}$
\item[] \textbf{Step 2.} Solve $\bx_{k+1} \leftarrow \argmax_{\bx \in \bX_{cand}} \hat{s}^2_{k}(\bx;\tau^2) / \hat{s}^2_{G \backslash (k \cup \bx)}(\bx;\tau^2)$
\item[] \textbf{Step 3.} Let $\bX_{k+1} \leftarrow \bX_{k} \cup \bx_{k+1}$, and $\bX_{cand} \leftarrow \bX_{cand} \backslash \bx_{k+1}$
\item[] \textbf{Step 4.} If $k+1=N$, then stop; otherwise let $k=k+1$ and go to step 1
\item[] \textbf{Output:} $\bX_{N}$
\end{itemize}
}} \\[10pt]
In step 2, a GP emulator is assigned to the set of points $D_k=(\bX_k,\by_k)$, and, for each $\bx \in \bX_{cand}$, a GP emulator is assigned to $\bX_G \backslash (\bX_k \cup \bx)$. The GP over $\bX_G \backslash (\bX_k \cup \bx)$ is required in order to estimate the difference between the total information and the information we have obtained by $\bX_k \cup \bx$.

Assuming the covariance is known, Krause et al. \cite{KSG2008} demonstrated that the MI algorithm for a sensor placement problem on an equidistant mesh can achieve a good accuracy at a relatively low computational cost compared to ALC.

\subsubsection{Example: A stationary Gaussian random field}
As a first example, we consider a realization of a stationary Gaussian random field with zero mean, and SE covariance function with $\sigma^2=1$ and $\sbxi=\left( 0.8,0.5 \right)^T$, on a $21\times21$ regular grid over $[0,1]^2$. The candidate set is a regular sub-grid of size $11\times11$. The remaining 320 design points are used to calculate the prediction accuracy. In all our numerical examples, the prediction accuracy is measured by the normalized RMSPE:
\begin{equation}
\label{eq:rmspe}
\operatorname{RMSPE}=\sqrt{\sum^{m}_{j=1} \dfrac{(y(\bx_j)-\hat{y}(\bx_j))^2}{m}},
\end{equation}
where the validation data set $\{\bx_j\}^m_{j=1}$ consists of $m$ input values at which the difference between the simulator output value $y(\bx_j)$ and the predicted value $\hat{y}(\bx_j)$ is evaluated. The \emph{normalized} RMSPE is given by $\operatorname{RMPSE}/(\max_{j}{y(\bx_j)}-\min_{j}{y(\bx_j)})$.

\begin{figure}[!ht]
\minipage{0.48\textwidth}
  \centering
  \includegraphics[width=0.70\linewidth]{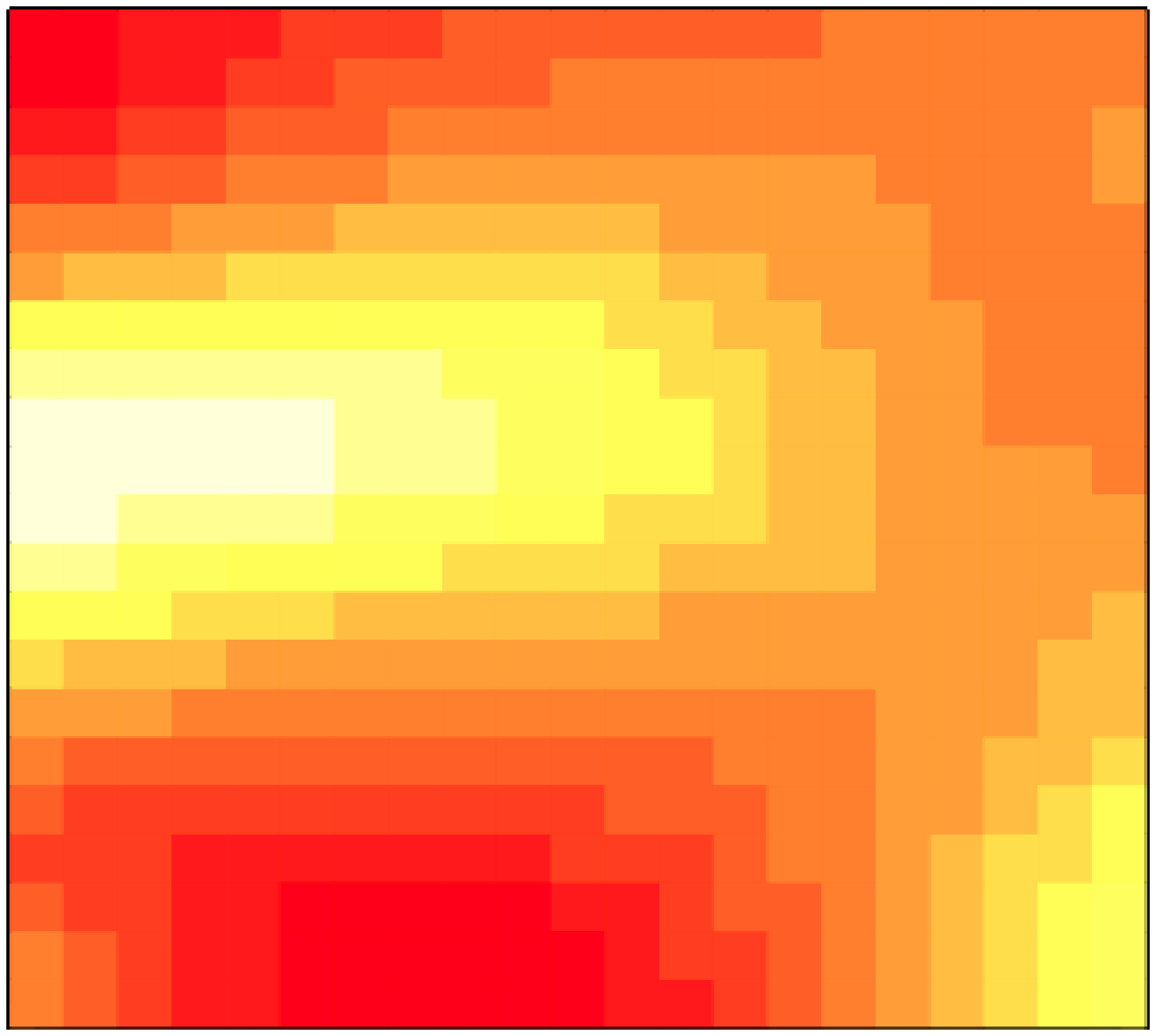}
\endminipage\hfill
\minipage{0.48\textwidth}
  \centering
  \includegraphics[width=0.8\linewidth]{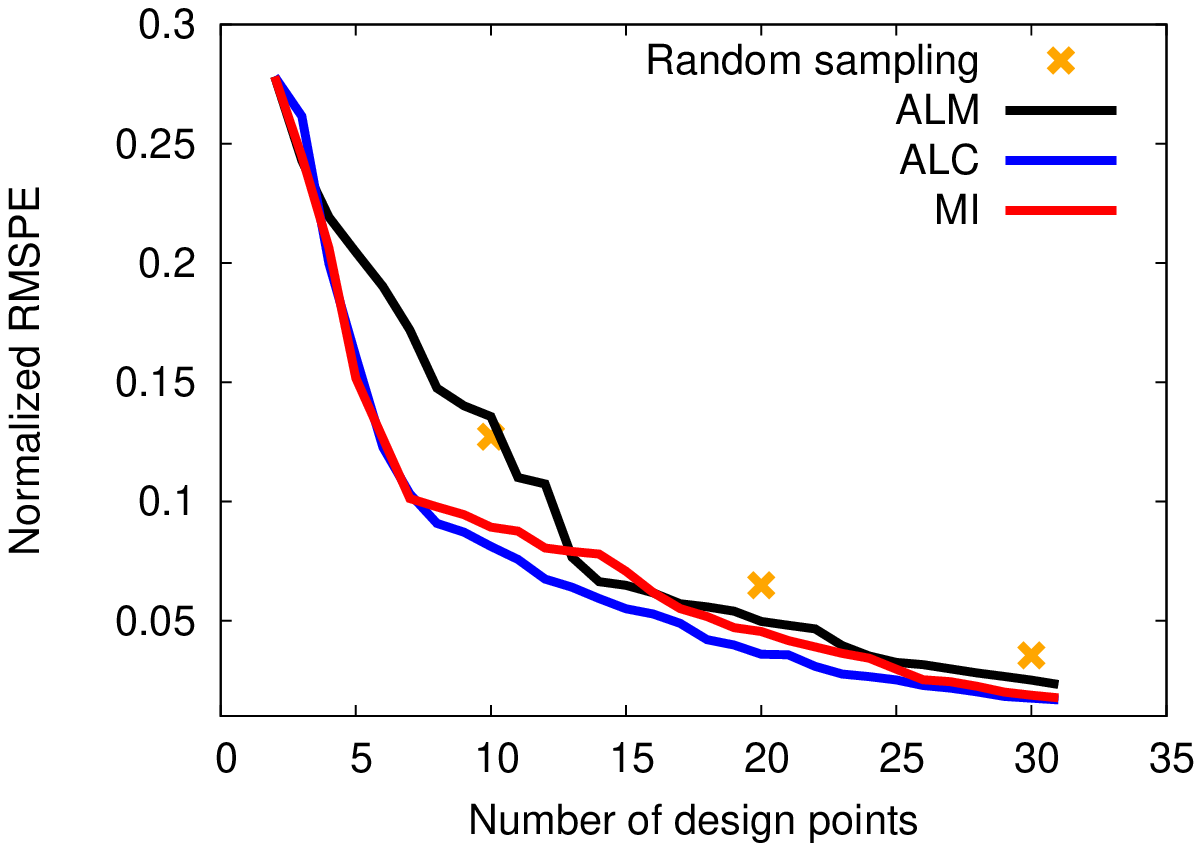}
\endminipage\hfill
\caption{Left: A realization of the stationary Gaussian random field. Right: Prediction errors.}\label{fig:grf2d}
\end{figure}

Figure \ref{fig:grf2d} shows results obtained for ALM, ALC and MI. The prediction errors are given as averages of ten tries with different initial two-point designs. The average performance of random sampling over 100 tries has also been included for comparison, and as expected, the sequential designs outperform random sampling. MI and ALC perform similarly. ALM performs the worst, partially because it systematically places most points on the boundary of the domain, and, as a result, not capturing well the large variation in the interior.

\subsubsection{A practical issue with the MI criterion}
\label{sec:issue}
Krause et al. \cite{KSG2008} showed theoretically and demonstrated empirically that the MI criterion is a promising criterion for sequential design of sensor networks on a discrete space. In computer experiments, however, the design space is generally not discrete but a compact subset of $\Rset^p$ where $p$ can be quite large. For the MI criterion to be considered, we have to discretize $\sX$ into a finite set $\bX_G \subset \sX$ of a grid $G$. This is because for each candidate point $\bx^* \in \bX_{cand}$, we want to assign a GP emulator over the points of a finite set $\bX_G \backslash (\bX_k \cup \bx^*)$ that approximates well $\sX \backslash (\bX_k \cup \bx^*)$. Recall that $\bX_k$ is the set of design points at stage $k$ of the sequential design.

We have observed that the MI criterion \eqref{eq:micriterionnew} is very sensitive to the distribution of points in $\bX_G$. For example when the points of $\bX_G$ are irregularly distributed, e.g., if some points are clustered, this criterion is unreliable. More specifically, if the criterion is evaluated at a point $\bx^* \in \bX_{cand}$ that is close to a point in $\bX_G \backslash (\bX_k \cup \bx^*)$, then the denominator $\hat{s}^2_{G \backslash (k \cup \bx^*)}(\bx^*)$ can become very small, and, as a result, producing a high MI score. In this situation, the issue is that the location of $\bx^*$ in relation to the current design $\bX_k$, which should be important factor, has little influence. This issue did not present itself in \cite{KSG2008}, since they considered an equidistant grid.

\begin{figure}[!ht]
\centering
\includegraphics[width=0.45\linewidth]{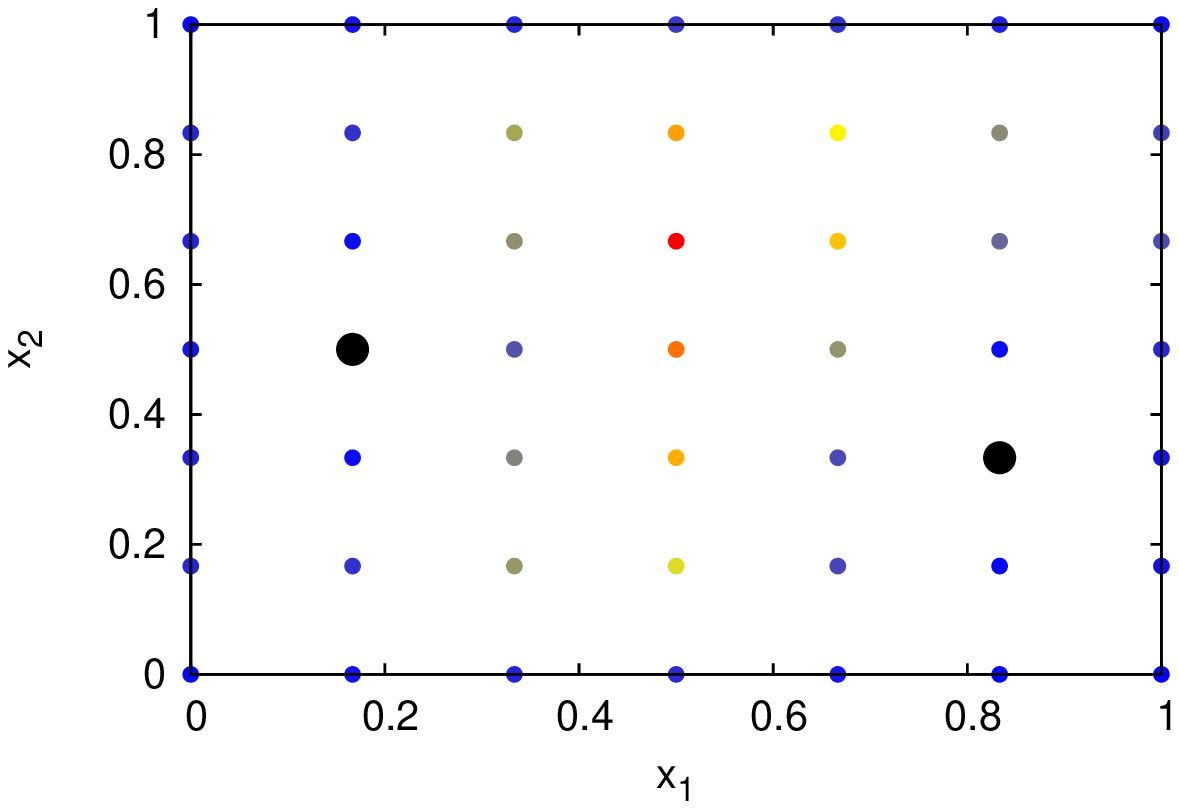}
\includegraphics[width=0.45\linewidth]{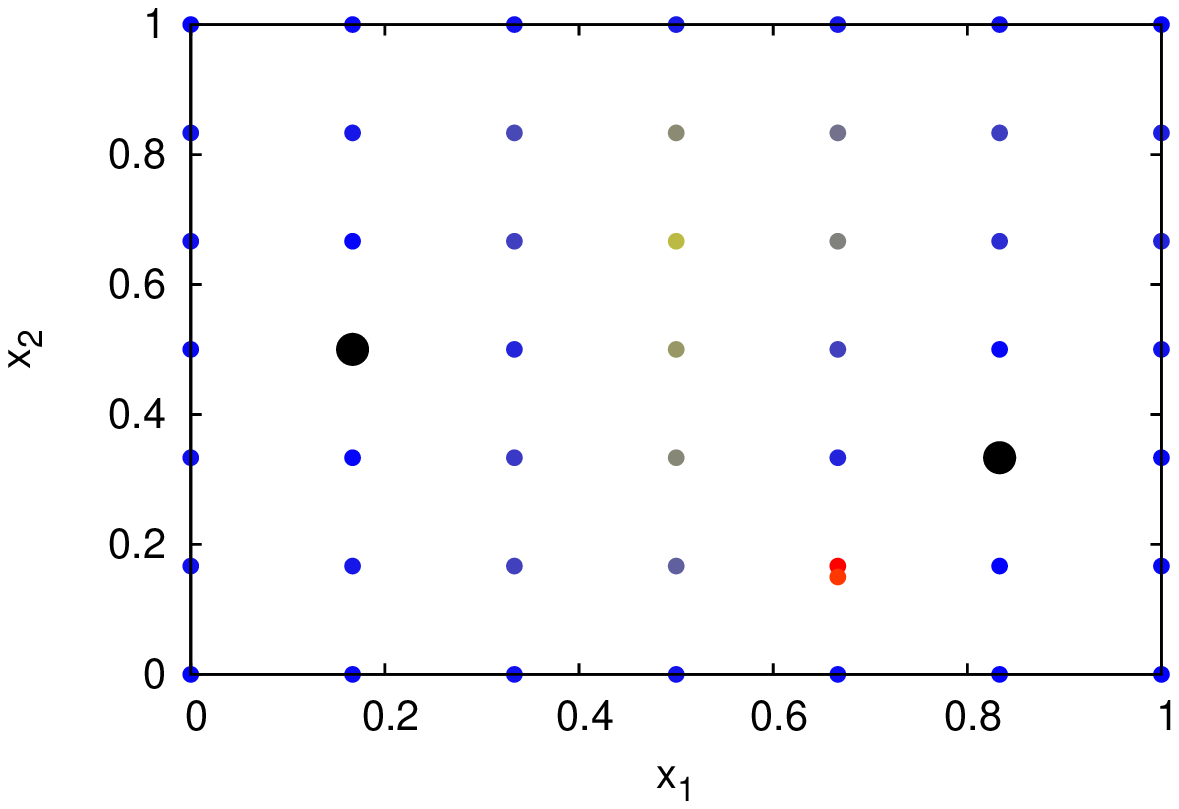}
\caption{The score value of the MI criterion over a $7 \times 7$ equidistant grid (Left), and of the same grid with an additional point at $(2/3,0.15)$ (Right).}\label{fig:issue2}
\end{figure}

See Figure \ref{fig:issue2} for an illustrative example where the MI criterion performs poorly. Two cases are considered: an equidistant grid $\bX_G$, and $\bX_G$ with an additional point $(2/3,0.15)$, that is, $\bX_G \cup \{ (2/3,0.15) \}$. A high MI score is marked in red, an intermediate score is yellow, and a low score is blue. The black dots are the points of design $\bX_k$.

\begin{figure}[!ht]
\centering
\includegraphics[width=0.45\linewidth]{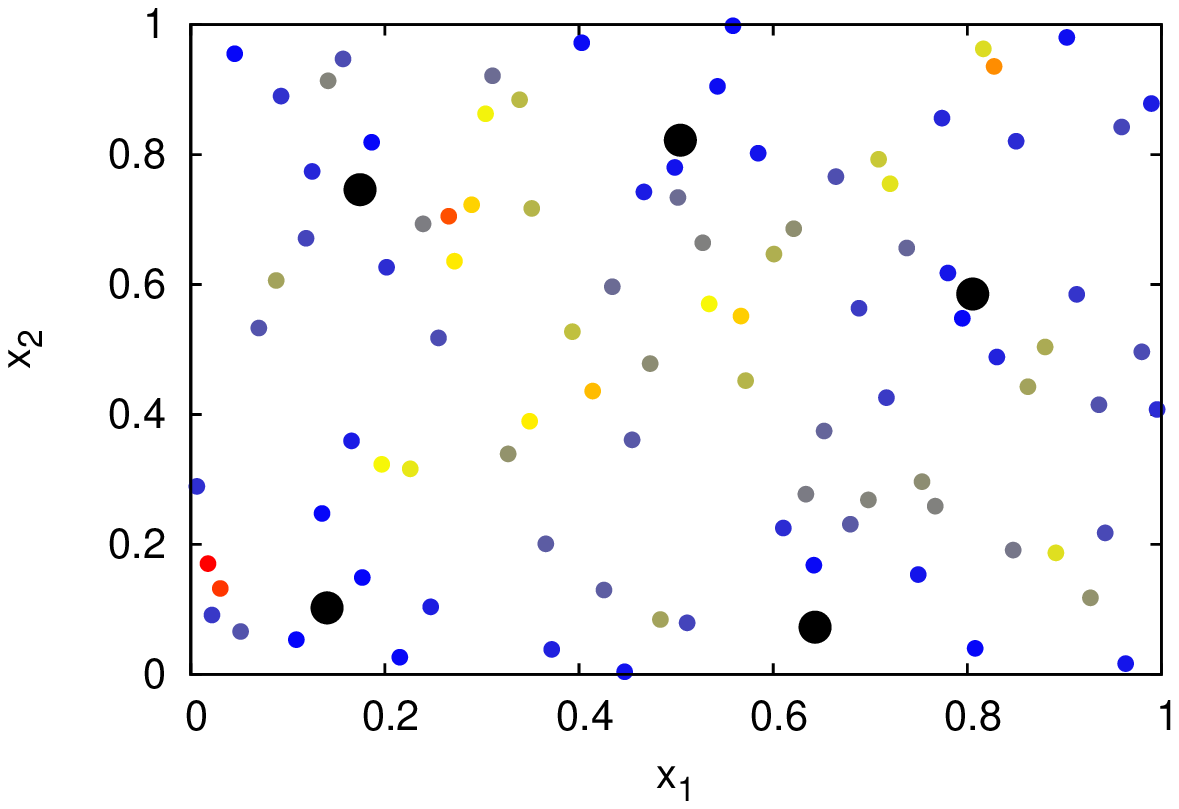}
\includegraphics[width=0.45\linewidth]{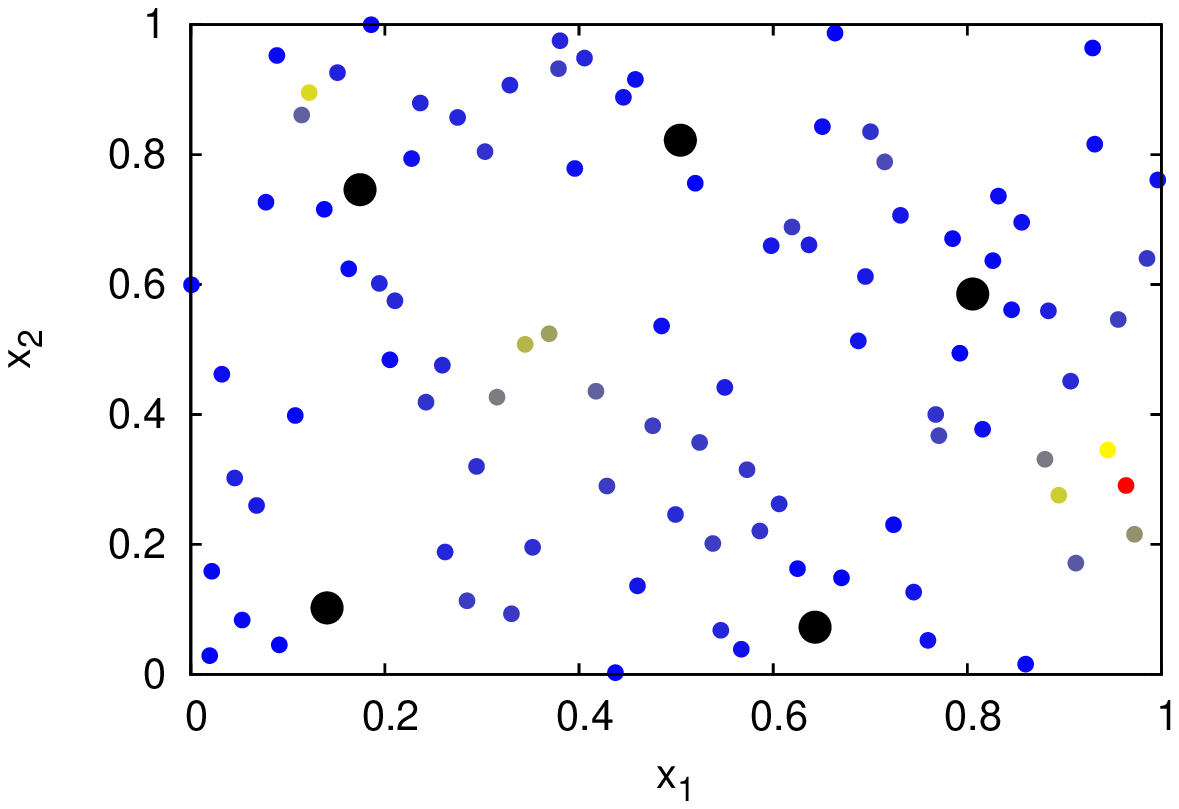}
\caption{Given a design $\bX_5$, the score value of the MI criterion is displayed for two maximin LHD candidate sets of size $100$.}\label{fig:issue2lhd}
\end{figure}

Two different choices of $\bX_G$ can result in highly conflicting MI scores, as demonstrated in Figure \ref{fig:issue2lhd} with two different maximin LHDs of size 100. Evidently, the MI criterion is not robust whenever the points are irregularly spaced. Moreover, for moderate to high dimensional spaces, typical of computer experiments, equidistant grids are too large to consider.

\subsection{Mutual information for computer experiments}
\label{sec:micealg}
In this section we present a sequential design algorithm called MICE (Mutual Information for Computer Experiments) for prediction. The algorithm uses on a modified MI criterion, and the correlation parameters are estimated adaptively using maximum likelihood. We also suggest that discretization $\bX_G$ of $\sX$ should not be held fixed, but instead a new $\bX_G$ should be sampled at each iteration.

\subsubsection{The MICE criterion: a modified MI criterion}
\label{sec:modmi}
We modify the MI criterion by introducing a parameter $\tau^2>0$ to the diagonal elements of the correlation matrix to smooth the prediction. Such a parameter is often called a \emph{nugget parameter}. $\bK_{\sbxi}$ is replaced by $\bK_{\sbxi,\tau^2}=\bK_{\sbxi}+\tau^2\Id$, where $\Id$ is the $n \times n$ identity matrix. A nugget parameter $\tau^2$ is commonly used to stabilize the inversion, using the Cholesky decomposition, of a possibly ill-conditioned correlation matrix. When $\tau^2$ is introduced to achieve numerical stability, it is usually chosen to be very small. Moreover, Gramacy and Lee \cite{GL2012} argue in favor of using a nugget parameter to smooth the prediction. The BLUP model, Eq. \eqref{eq:predmean}, with a non-zero nugget is not a perfect interpolator of the data, and our Theorem \ref{theo:nugget} below clarifies the impact a nugget parameter has on the GP emulator variance for any point in $\bX$. Clearly, if $\tau^2=0$, $\hat{s}^2(\bx_i)=0$ for $\bx_i \in \bX$.

\begin{theorem}\label{theo:nugget}
For a GP emulator with constant mean on $(\bX,\by)$, the predictive variance, Eq. \eqref{eq:predvar}, at any design point $\bx_i \in \bX$ can be written as
\begin{equation}
\label{eq:nugget}
\hat{s}^2_{\tau^2}(\bx_i) = \sigma^2\left(\tau^2-\tau^4\textbf{e}_i^T(\bK+\tau^2\Id)^{-1}\textbf{e}_i+\tau^4\dfrac{(\textbf{e}_i^T(\bK+\tau^2\Id)^{-1}\textbf{1})^2}{\textbf{1}^T(\bK+\tau^2\Id)^{-1}\textbf{1}}\right),
\end{equation}
where $\tau^2>0$ is a nugget parameter, and $\textbf{e}_i$ is the $i$-th unit vector.
\end{theorem}

According to Theorem \ref{theo:nugget}, whenever $\tau^2>0$ is added to the correlation matrix diagonal, the variance of a GP emulator at a design point consists of terms in the order of $\sigma^2\tau^2$ and $\sigma^2\tau^4$. In practice, the nugget $\tau^2$ is usually orders of magnitude smaller than 1. In Eq. \eqref{eq:nugget}, the magnitude of the last term in the round brackets tends to be much smaller than the other two; hence, the predictive variance is here typically smaller than $\sigma^2\tau^2$. Moreover, as $\tau^2$ increases, the second and third term approaches $\tau^2$ and $\tau^2/k$, respectively, where $k$ is the number of points in the design. This follows from that as $\tau^2$ increases the inverse matrix reduces to $(\bK+\tau^2\Id)^{-1}\approx\tau^{-2}\Id$. Hence, if $\tau^2$ is large enough, we can show by a simple calculation using Theorem \ref{theo:nugget} that $\hat{s}^2_{\tau^2}(\bx_i) \approx \sigma^2\tau^2/k$ for $\bx_i \in \bX_k$. 

\begin{figure}[!ht]
\centering
\includegraphics[width=0.6\linewidth]{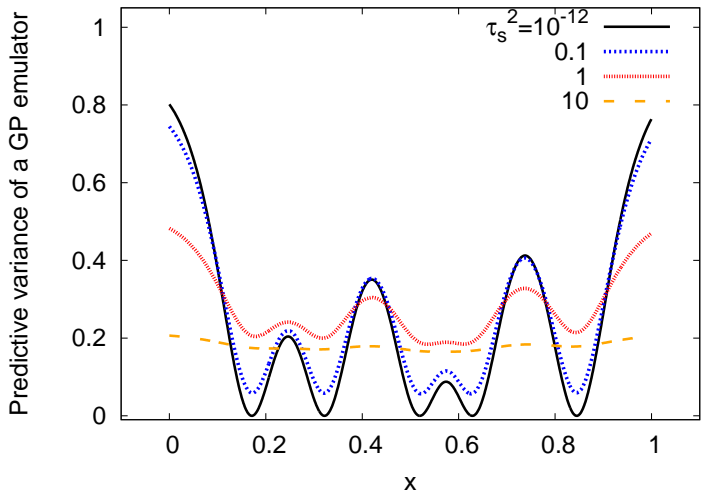}
\caption{The predictive variance of a GP emulator as a function of $\tau_s^2$ for a one-dimensional problem in domain $[0,1]$.}\label{fig:predvarGP}
\end{figure}

In the sequential design, we define the MICE criterion as follows:
\begin{equation}
\label{eq:micecriterion}
\bx_{k+1}=\argmax_{\bx \in \bX_{cand}} \hat{s}^2_{k}(\bx)/\hat{s}^2_{G \backslash (k \cup \bx)}(\bx;\tau^2_s),
\end{equation}
where a nugget parameter $\tau_s^2>0$ (s for smoothing) is added to the correlation matrix $\bK$ of the GP on $\bX_G \backslash (\bX_k \cup \bx)$  (in the denominator) with the specific purpose of flattening its variance. The flattening of the variance is performed as a means of preventing the denominator term to be close to zero, which may happen whenever a candidate point $\bx^*$ is too close to a point in $\bX_G \backslash (\bX_k \cup \bx^*)$. Figure \ref{fig:predvarGP} shows the predictive variance for different choices of $\tau^2_s$. 

\begin{figure}[!ht]
\centering
\includegraphics[width=0.6\linewidth]{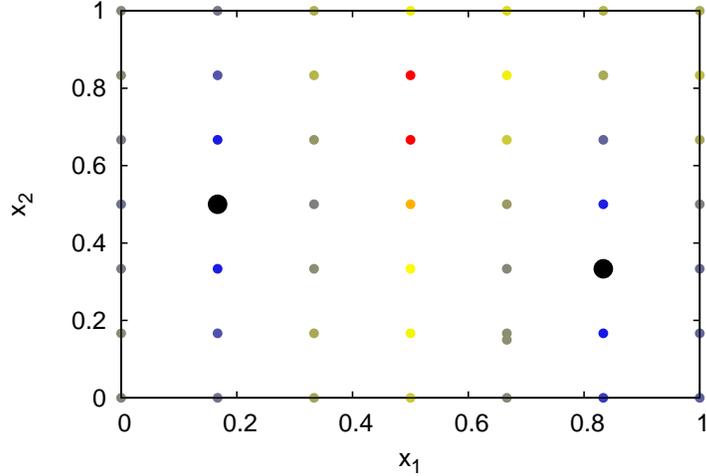}
\caption{The score values of the MICE criterion using $\tau^2_s=1$ for a $7 \times 7$ equidistant grid with an additional candidate point at $(2/3,0.15)$.}\label{fig:issue2mice}
\end{figure}

\begin{figure}[!ht]
\centering
\includegraphics[width=0.45\linewidth]{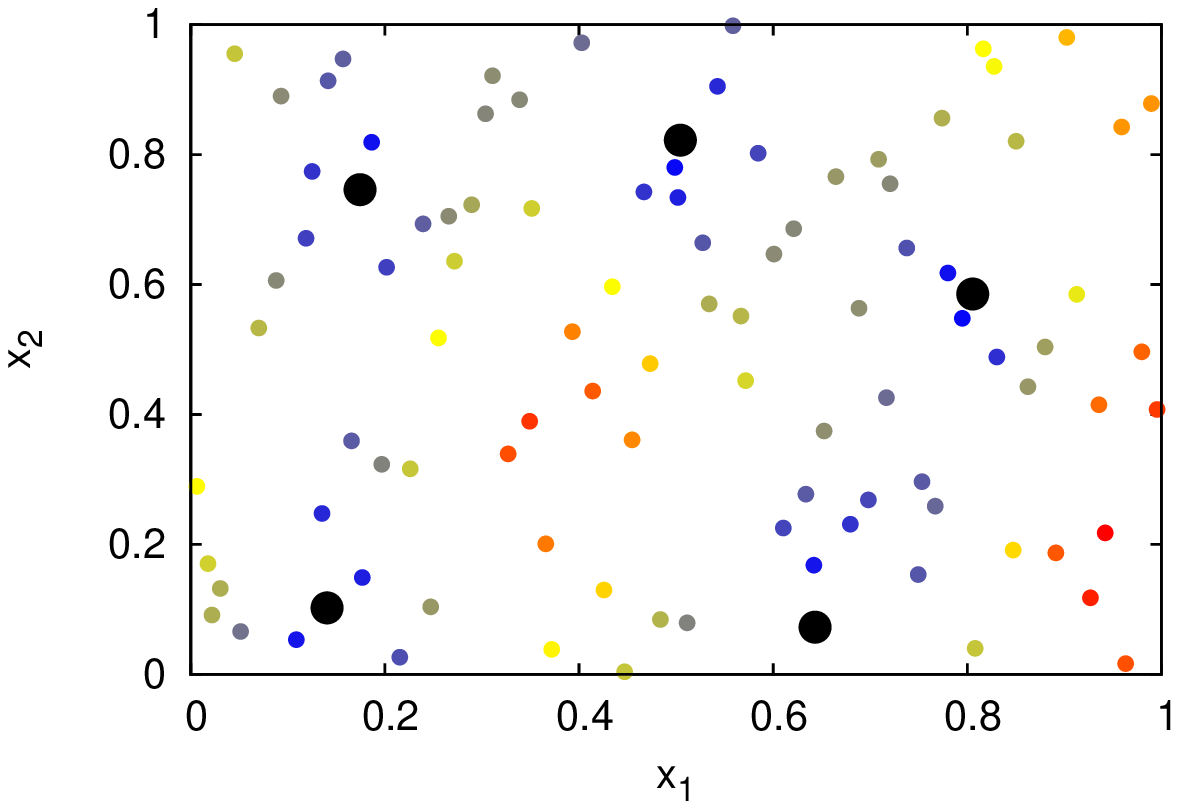}
\includegraphics[width=0.45\linewidth]{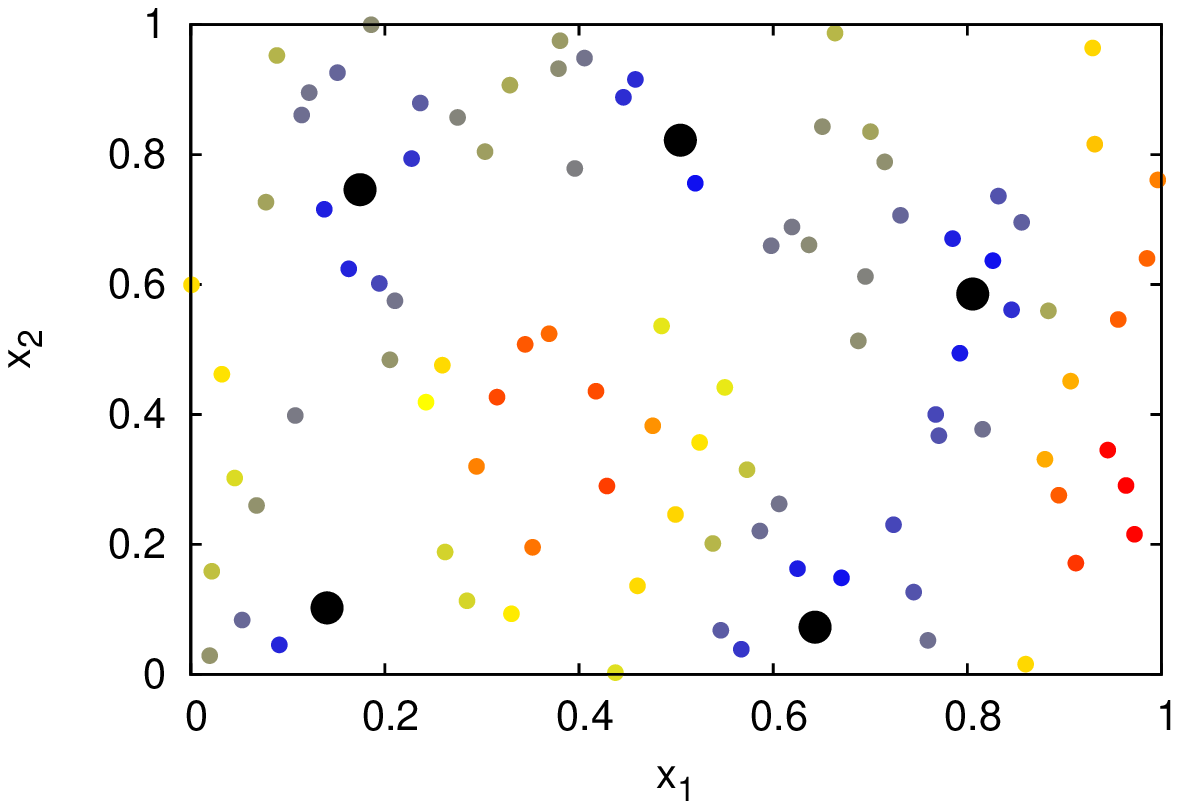}
\caption{Given a design $\bX_5$, the score values of the MICE criterion using $\tau^2_s=1$ are shown for two maximin LHD candidate sets of size $100$.}\label{fig:issue2lhdmice}
\end{figure}

The sweet spot of $\tau^2_s$ is around $1$, where the variance is not close to zero and the shape of the variance curve is well preserved. Hence our default choice is $\tau^2_s=1$. Figure \ref{fig:issue2mice} and \ref{fig:issue2lhdmice} show MICE scores with $\tau^2_s=1$, which can be compared with the corresponding figures for MI (see Figure \ref{fig:issue2} and \ref{fig:issue2lhd}, respectively). By examining the figures, we can conclude that MICE is more robust than MI. For a simple regular grid, MICE and MI perform the same.


\subsubsection{Adaptivity}
The original implementation of the MI algorithm assumed that the covariance is fully known, but that is rarely the case in modeling of computer experiments. Therefore, in our implementation, whenever the correlation parameters $\sbxi$ are unknown, we provide point estimates that maximize the likelihood. This approach is described in Section \ref{sec:uncertain}. The MLEs of $\sbxi$ are sequentially updated at each stage $k$, denoted by $\hat{\sbxi}_k$, by using all available input-output data. However, the updating may be skipped at some stages in order to make computational savings.

\begin{figure}[!ht]
\centering
\includegraphics[width=0.6\linewidth]{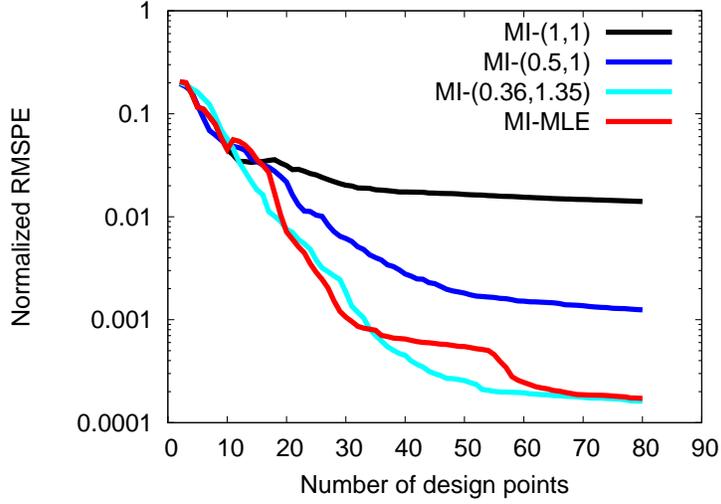}
\caption{Prediction errors for the MI algorithm when using estimates of $\sbxi$ (three guesses, and one using MLE updates).}\label{fig:issue3}
\end{figure}

The prediction errors for different choices of estimates for the correlation parameters are shown in Figure \ref{fig:issue3}, where the example is the so-called Branin function, $y(\bx)=(x_2-5.1x_1^2/(4\pi^2)+(5/\pi)x_1-6)^2+10(1-1/(8\pi))\cos(x_1)+10$, on a $21 \times 21$ regular grid over $[-5,10]\times[0,15] \subset \Rset^2$. A GP emulator is used with the Mat\'ern correlation fixed at $\nu=5/2$. Here, MI-MLE is the MI algorithm with the addition of a MLE step at each stage $k$ of the sequential design. For $k<10$, the tentative values $(1,1)$ are assigned for $\sbxi$. Three fixed guesses of $\sbxi$ are considered: $(1,1)$, $(0.5,1)$, and $(0.36,1.35)$. The latter guess is the final MLEs obtained by MI-MLE. The results show the importance of having good estimates of the correlation parameters, and that the MLE method can greatly improve upon simple guesses.

\subsubsection{The MICE algorithm}
The \emph{MICE algorithm} is outlined below with some details on some of the steps. \\

\textbf{MICE algorithm:} \\
\fbox{\parbox{0.95\textwidth}{
\begin{itemize}
\item[] \textbf{Require:} Function $y(\bx)$, GP emulator ($\bh(\cdot)$,$K(\cdot,\cdot;\sbxi)$), nugget parameters $\tau^2$ and $\tau^2_s$, design space $\sX$, initial data $(\bX_k,\by_k)$, discrete set size $N_G$, candidate set size $N_{cand}$, desired design size $N$
\item[] \textbf{Step 1.} MLE to obtain estimates $\hat{\sbxi}_k$ of $\sbxi$ in $K(\cdot,\cdot;\hat{\sbxi}_k)$
\item[] \textbf{Step 2.} Fit GP emulator to data $(\bX_k,\by_k)$
\item[] \textbf{Step 3.} Generate a discrete set $\bX_{G}$ of size $N_G$, and choose a candidate set $\bX_{cand} \subseteq \bX_{G}$
\item[] \textbf{Step 4.} Solve $\bx_{k+1}=\argmax_{\bx \in \bX_{cand}} \hat{s}^2_{k}(\bx;\hat{\sbxi}_k,\tau^2)/\hat{s}^2_{G \backslash (k \cup \bx)}(\bx;\hat{\sbxi}_k,\max\{\tau^2,\tau_s^2\})$
\item[] \textbf{Step 5.} Evaluate $y_{k+1}=y(\bx_{k+1})$, and let $\bX_{k+1}=\bX_k \cup \bx_{k+1}$ and $\by_{k+1}=\by_{k} \cup y_{k+1}$
\item[] \textbf{Step 6.} If $k+1=N$, then stop; otherwise let $k=k+1$, and go to step 1
\item[] \textbf{Output:} $D_N=(\bX_{N}$,$\by_{N})$ of size $N$
\end{itemize}}} \\[10pt]
In step 3, we suggest that $\bX_G$ is sampled in the design space $\sX$, instead of keeping $\bX_G$ fixed throughout. In our examples, the size of $\bX_G$ is $k+N_G$, where $k$ is the number of points of $X_k$. The additional $N_G$ points are generated picking a LHD from a set of LHDs by maximizing the minimum distance between the points in this LHD and the current design $\bX_k$. In step 4, the MICE criterion is evaluated for all $\bx \in \bX_{cand}$. The choice of $\tau^2_s$ is critical; more on this in Section \ref{sec:5}. Note that the parameter $\tau^2_s$ is introduced to the GP for design $\bX_{G} \backslash (\bX_k \cup \bx)$, and not to the GP for design $\bX_k$. Although, a nugget parameter $\tau^2>0$ can still be introduced to any GP for other purposes such as achieving numerical stability (typically much smaller than $\tau^2_s=1$). We assume that the correlation parameters are the same for the GP on $\bX_G \cup (\bX_k \cup \bx)$ as for the GP on $\bX_k$.

\subsubsection{Near optimality results}
We here provide an approximative bound of optimality for the MICE algorithm based on near optimality results in \cite{KSG2008} for the MI algorithm under known $\sbxi$.  More generally, our results account for the possibility that a different nugget parameter is used for the GP over $\bX_k$ than the over $\bX_\bG \backslash \bX_k \cup \bx$.

\begin{theorem}\label{theo:opt} 
Let $Y(\bx)$ be a second-order stationary Gaussian process with constant mean on a compact set $\sX \subset \Rset^p$ with a continuous correlation function $K(\bx,\bx'): \sX \times \sX \to \Rset_0^{+}$. Assume that we have estimates $\hat{\sbxi}_k$ for $\sbxi$ at stage $k$ that satisfy for some constant $\alpha>0$, $\vert K(\bx,\bx';\sbxi)-K(\bx,\bx';\hat{\sbxi}_k) \vert \le \alpha$. Then, for any $\varepsilon>0$, and any finite number $N$, there exists a discretization $\bX_{G}$ of mesh width $\delta>0$ such that MICE is guaranteed to select a design $D_N=(\bX_N,\by_N)$ with $N$ design points, where $N\le 2\vert \bG \vert$, for which
\begin{equation*}
MI(D_N)\ge(1-1/e)(OPT-N\varepsilon-2(\alpha\sigma^{-1}\tau^{-1})^2N^4(1+N^{3/2})^2-N^{3}\sqrt{N}\vert \tau_s^2-\tau^2\vert/\tau^2_s),
\end{equation*}
where $e$ is the base of the natural logarithm, $OPT$ is the value of the mutual information for the optimal design of size $N$, and, $\tau^2$ and $\tau^2_s$ are nugget parameters in the correlation matrices for $\bX_{k}$, and $\bX_G \backslash \bX_{k}$, respectively.
\end{theorem}

Under perfect conditions the upper bound in Theorem \ref{theo:opt} guarantees a performance within 63$\%$ of the optimum. The term $N\varepsilon>0$ is essentially zero as long as the discretization $\bX_G$ is fine enough. The term $2(\alpha\sigma^{-1}\tau^{-1})^2N^4(1+N^{3/2})^2$ is non-zero in the presence of parameter uncertainty, and the term $N^{3}\sqrt{N}\vert \tau_s^2-\tau^2\vert/\tau^2_s$ appears when a nugget $\tau^2$ is used for the GP emulator over $\bX_k$. Our extension of the approximative bound of optimality to MICE reveals the effect of $\tau^2_s$ on the performance. Our default choice $\tau^2_s=1$ is not causing the algorithm to diverge too much from MI, as long as $\tau^2$ is not much larger than $\tau^2_s$. In addition, whenever the correlation parameters are poorly estimated, the optimality bound is not sharp. To increase our understanding of the MICE behavior with respect to the choice of $\bX_G$, we provide the following theorem:

\begin{theorem}\label{theo:bound}
Let $Y(\bx)$ be a second-order stationary Gaussian process with constant mean on a compact subset $\sX$ of $\Rset^p$ with a Lipschitz-continuous correlation function. Then, for any $\varepsilon>0$, there exists a regular grid $\bX_{G} \subset \sX$ with grid spacing $\delta=2\varepsilon/(\sqrt{p}K_{L})$ so that for any untried point $\bx^* \in \sX$ the predictive variance $\hat{s}^2_{\tau^2}(\bx)$ is bounded as
$$
-\tau^4b_1(\tau^2)-\varepsilon<\sigma^{-2}\hat{s}^2_{\tau^2}(\bx^*)-\tau^2<\tau^4b_2(\tau^2)+\varepsilon,
$$
where 
$$
b_1(\tau^2)=\max\left\{ \textbf{e}_i^T(\bK+\tau^2\Id)^{-1}\textbf{e}_i : \bx_i \in \bX_{G} \right\},
$$
and
$$
b_2(\tau^2)=\max\left\{ \dfrac{(\textbf{e}_i^T(\bK+\tau^2\Id)^{-1}\textbf{1})^2}{\textbf{1}^T(\bK+\tau^2\Id)^{-1}\textbf{1}} : \bx_i \in \bX_{G} \right\},
$$
where $\Id$ is the identity matrix, and $\textbf{e}_n$ the $i$-th unit vector for member $\bx_i$ of $\bX_{G}$. Here $K_L$ is the Lipschitz constant for $\hat{s}^2_{\tau^2}(\bx)$ over $\sX$.
\end{theorem}

Theorem \ref{theo:bound} tells us that when $\bX_G$ is a regular grid dense enough in $\sX$, while $\tau^2$ and $\tau_s^2$ are small enough, MICE is equivalent to ALM. In fact, MICE also behaves as ALM if $\bX_G$ more dense, and $\tau_s^2$ large enough so that $(\bK+\tau^2\Id)^{-1}=\tau^{-2}\Id$ (approximately), since according to Theorem \ref{theo:bound}, as $\varepsilon>0$ becomes arbitrary small, then $0<\hat{s}^2_{G \backslash (k \cup \bx)}(\bx)<\varepsilon$. This can be seen in Figure \ref{fig:predvarGP}. Nonetheless, with $\tau_s^2=1$, MICE is not expected to behave as ALM. Similarly, MI behaves as ALM whenever $\bX_{G}$ is dense in $\sX$, and $\tau^2$ is very small. The prerequisites of Theorem \ref{theo:bound} hold in our numerical tests, because both the SE correlation function and Mat\'ern correlation with $\nu=5/2$ are continuously differentiable (hence Lipschitz continuous) \cite{HS1993}. 

\subsubsection{A computational improvement}

In MICE, we compute $\hat{s}_{G \backslash (k \cup \bx)}^2(\bx)$ for all $\bx \in \bX_{cand}$, which requires the Cholesky decomposition of a $(N_{G}-k-1) \times (N_{G}-k-1)$ correlation matrix $\bK_{G \backslash (k \cup \bx)}$, where $N_G$ is the number of points in $\bX_G$. This is a computationally intensive task if $N_G$ is large. To overcome this, we use the following implementation which only requires a single Cholesky decomposition. First, invert the correlation matrix $\bK_{G \backslash k}$. Then, exploit the partitioned inverse formula for matrices in block-form. That is, the inverse of
\begin{equation} \bK_{G \backslash k} = \left( \begin{array}{cc}
\bK_* & \bk^*(\bx) \\
\bk^T_{*}(\bx)^T & K(\bx,\bx) \end{array} \right)
\end{equation}
can be written as:
\begin{equation}\label{eq:invpart} \bK_{G \backslash k}^{-1} = \left( \begin{array}{cc}
\bB & \bb_{12} \\
\bb_{21} & b \end{array} \right)
\end{equation}
where $\bK_{*}=\bK_{G \backslash (k \cup \bx)}$, and $\bk_{*}(\bx)=\bk_{G \backslash (k \cup \bx)}(\bx)$. Here $\bB=\bK_{*}^{-1}+\frac{1}{k}\bK_{*}^{-1}\bk_{*}(\bx)\bk_{*}^T(\bx)\bK_{*}^{-1}$, $\bb_{12}=-\frac{1}{k}\bK_{*}^{-1}\bk_{*}(\bx)$, $\bb_{21}=-\frac{1}{k}\bk_{*}^T(\bx)\bK_{*}^{-1}$, and $b=1/(K(\bx,\bx)-\bk_{*}^T(\bx)^T\bK_{*}^{-1}\bk_{*}(\bx))$. This relates $\bK_{G \backslash (k \cup \bx)}^{-1}$ to $\bK_{G \backslash k}^{-1}$ for any $\bx \in \bX_{G \backslash k}$ as follows: given $\bK_{G \backslash k}^{-1}$, we can obtain $\bB$, $\bb_{12}$, $\bb_{21}$ and $b$, directly from Eq. \eqref{eq:invpart}, and then we find that $\bK_{G \backslash (k \cup \bx)}^{-1}=\bB-\frac{1}{b}\bb_{12}\bb_{21}$. Therefore, $\bK_{G \backslash (k \cup \bx)}^{-1}$ can be obtained from $\bK_{G \backslash k}^{-1}$ in $\bigO((N_{G}-k)^2)$.

\subsubsection{Example: a visualization of the design selection}
\begin{figure}[!ht]
\centering
\minipage{0.5\textwidth}
  \includegraphics[width=0.8\linewidth]{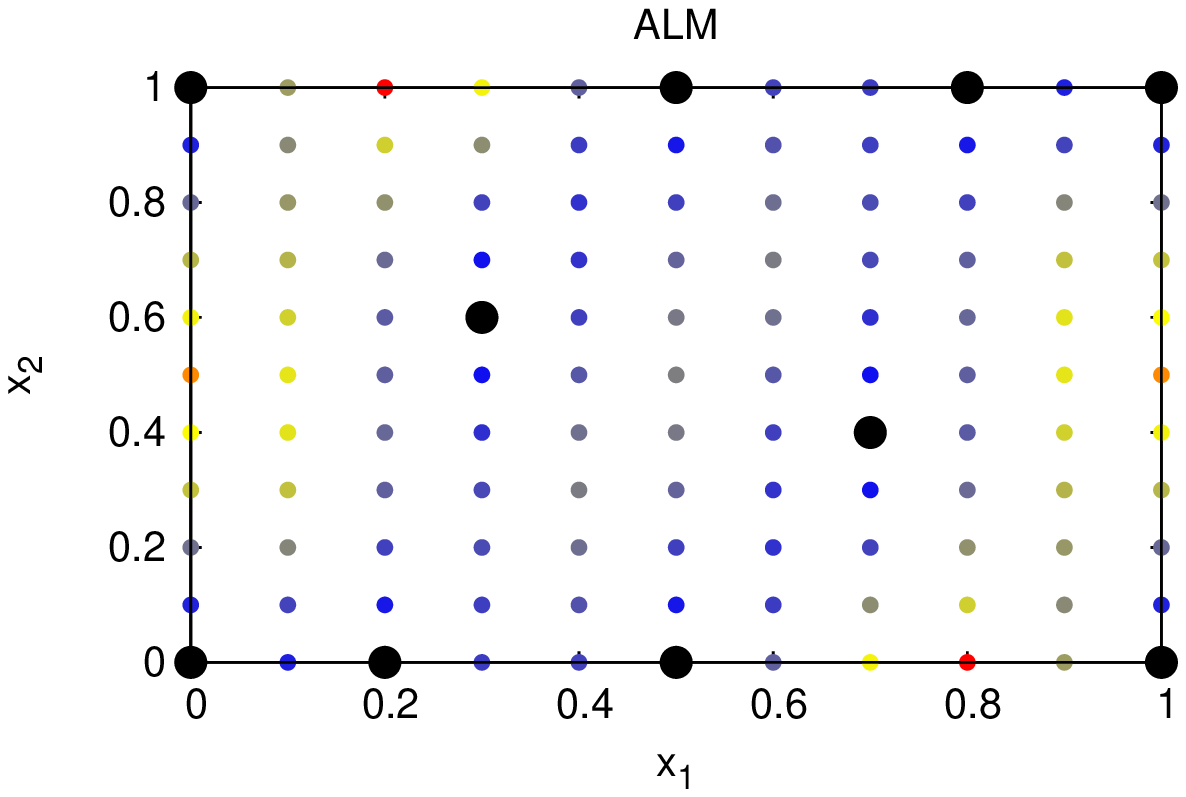}
  \includegraphics[width=0.8\linewidth]{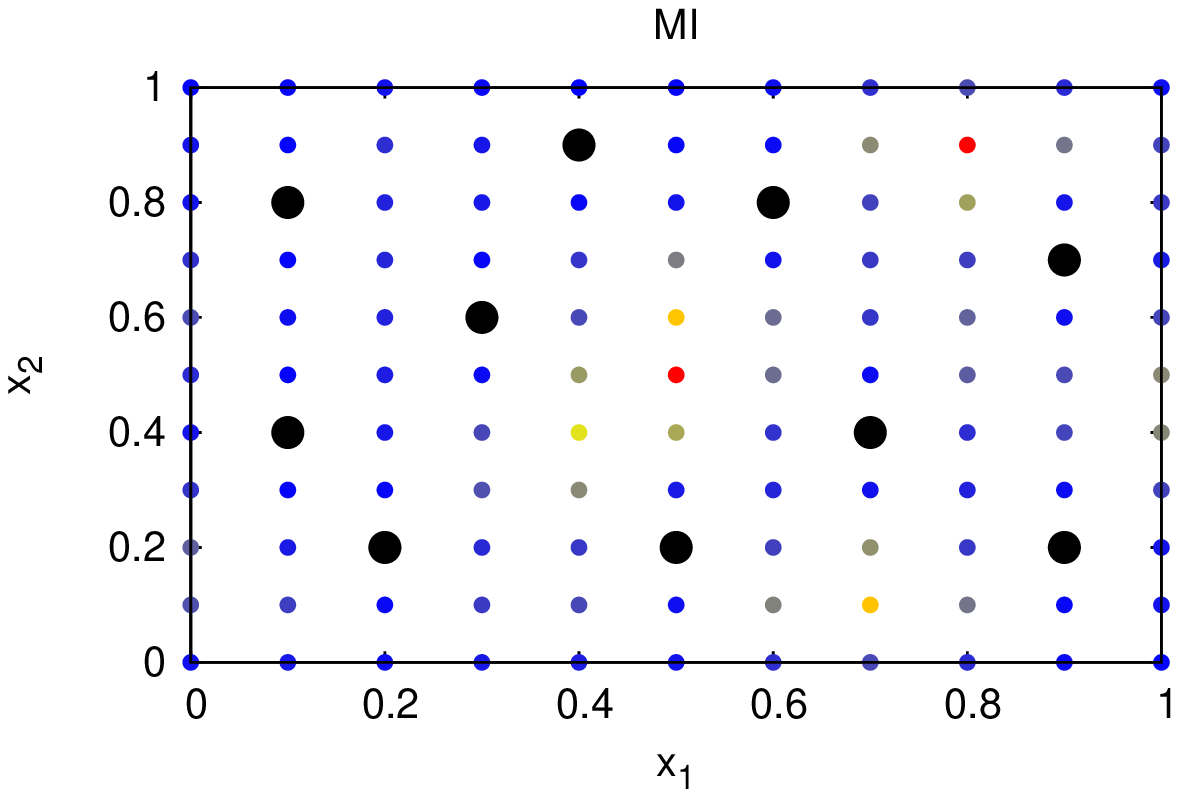}
\endminipage\hfill
\minipage{0.5\textwidth}%
  \includegraphics[width=0.8\linewidth]{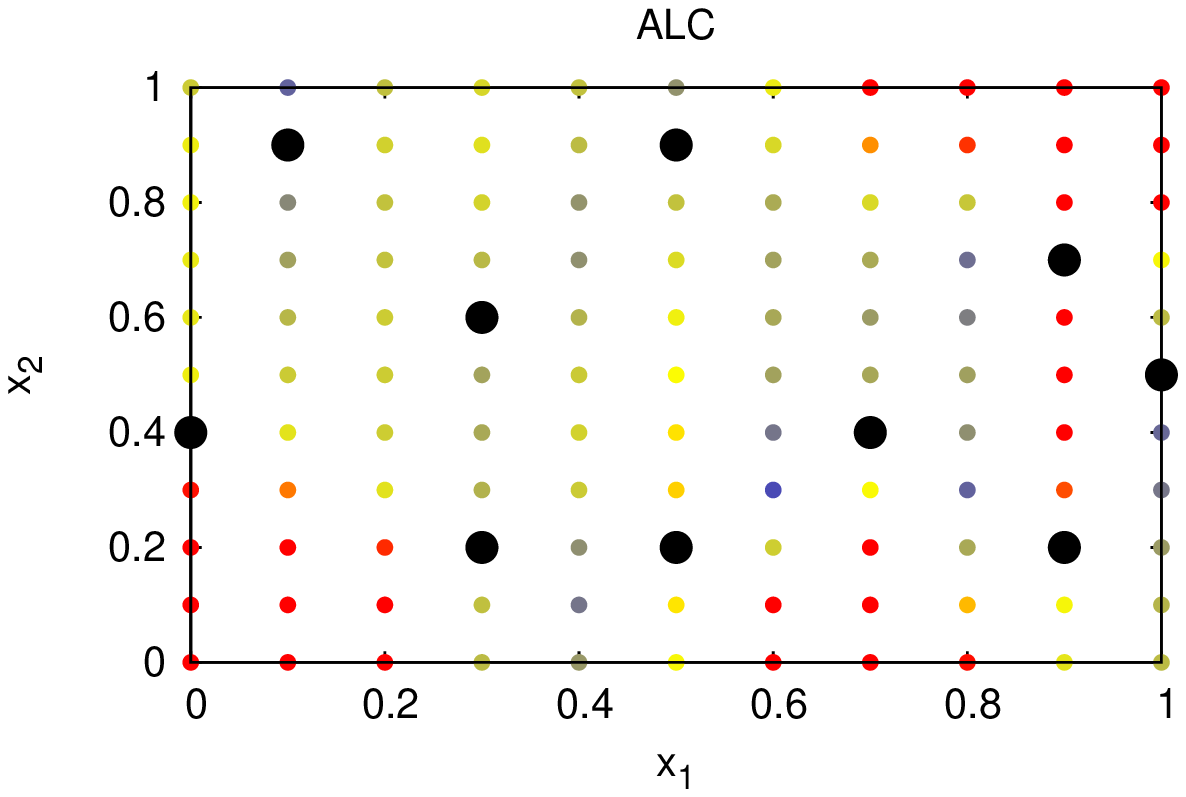}
  \includegraphics[width=0.8\linewidth]{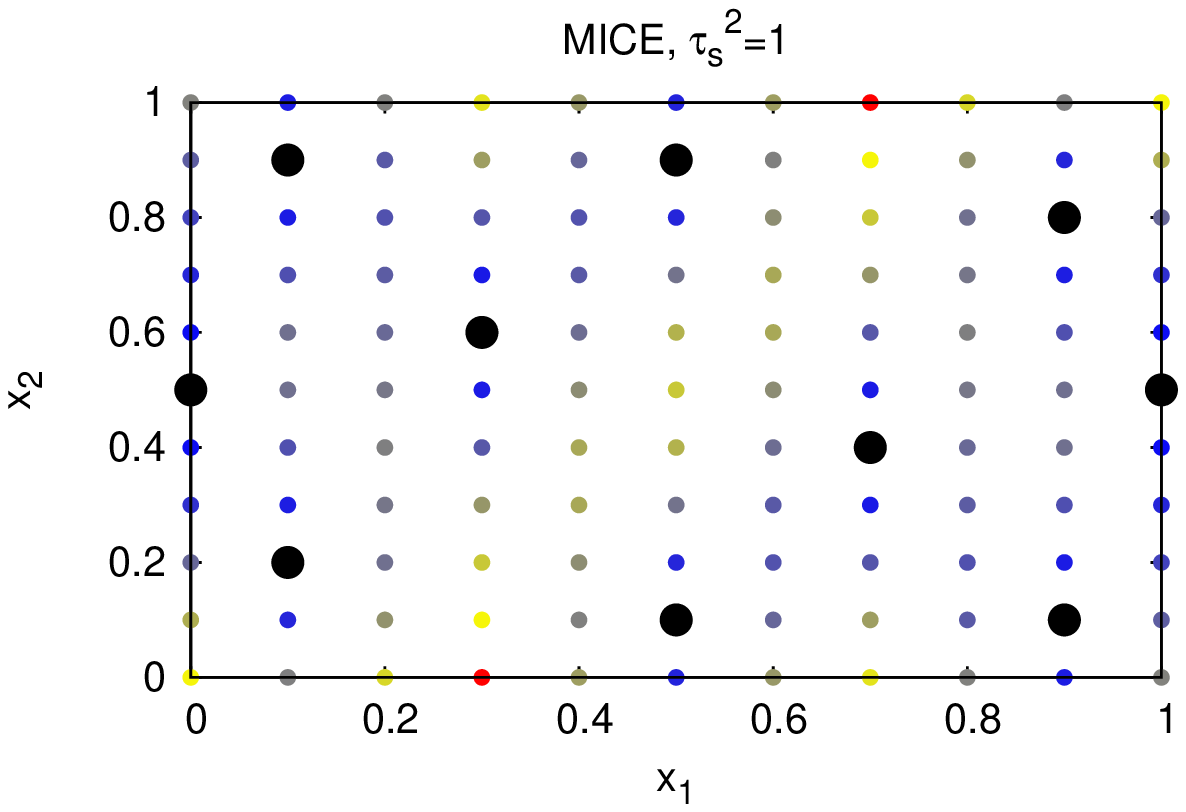}
\endminipage
\caption{Design selection with ALM, ALC, MI and MICE on $[0,1]^2$ }\label{fig:mice2D}
\end{figure}

Design selection with ALM, ALC, MI, and MICE, on $[0,1]^2$ are shown in Figure \ref{fig:mice2D}. A GP emulator with a constant mean is used with a fixed Mat\'ern covariance using $\sigma^2=1$, $\nu=5/2$ and $\sbxi=(0.4,1)$. The black-solid dots are design points, and the others are candidate points with the color representing the score value (red-high, blue-low) for the different design criteria. The initial design consists of the points $(0.3,0.6)$ and $(0.7,0.4)$. MICE with $\tau^2_s=1$ and ALC produce centered and well-spaced designs, whereas ALM focuses on the boundary. MI is the criterion most reluctant to select boundary points.

\section{Computational complexity}
\label{sec:4}
For the sequential algorithms the total computational cost to obtain a design $\bX_N$ of size $N$ may be divided into the cost to fit the GP emulator (in our case using MLE), the cost to generate the candidate set $\bX_{cand}$, and the cost to evaluate a design criterion over the candidate points:
\begin{equation}
T_{total}=T_{mle}+T_{cand}+T_{select}.
\label{eq:costformula}
\end{equation}
The time complexity to compute the MLEs is in $\bigO$($pN^2+N_{mle}N^{\omega}$), where $\omega>0$ is related to the efficiency of the algorithm for matrix inversion: for na\"ive Gaussian elimination $\omega=3$, and for Strassen's algorithm $\omega=\log_{2}{\left(7\right)}$. The term $pN^2$ is the number of operations needed to determine the distances between distinct pairs of points in $\bX$ (which is $N(N-1)/2$). The second term, $N_{mle}N^{\omega}$, is the time complexity of MLE, which is directly related to the cost of inverting the correlation matrix $\bK_{\sbxi}$, $N_{mle}$ times. $N_{mle}$ is the number of trial points visited during the optimization to find the MLEs of $\sbxi$. To train the GP emulator, that is to say, determine the weights $\sbl$ of the corresponding BLUP model \eqref{eq:predmean}, only matrix multiplications (each of order $\bigO\left( N^2 \right)$) are required. The time to evaluate the mean of the GP emulator at an untried point is $\bigO\left( pN \right)$, and to evaluate the variance is $\bigO\left( pN^2 \right)$. 

\begin{table}[!ht]
\centering
\caption{Time complexity for ALM, ALC and MICE.}
\label{tab:complexity}
\begin{tabular}{|l|l|}
\hline
Algorithm & Total time complexity for design size $N$ \\
\hline
ALM & $\bigO\left(N_{mle}N^{1+\omega}+N_{cand}pN^2\right)$ \\
ALC & $\bigO\left(N_{mle}N^{1+\omega}+N_{cand}N_{ref}pN^3\right)$ \\
MICE & $\bigO\left(N_{mle}N^{1+\omega}+N(N_G-N)^{\omega}+N_{cand}pN^3+N_{cand}Np(N_G-N)^{2}\right)$ \\
\hline
\end{tabular}
\end{table}

The computational complexity for the different algorithms is presented in Table \ref{tab:complexity}. For ALC, we have adopted formulation \eqref{eq:alcfinal}, which is the formulation with lowest computational cost. The time complexity for a single ALM step is $\bigO\left(pk^2+N_{mle}k^{\omega}+N_{cand}pk\right)$, where $k$ is the current design size. The total cost for ALM is $\bigO\left(pN^3+N_{mle}N^{1+\omega}+N_{cand}pN^2\right)$, where $N$ is the final design size. $N_{ref}$ is specific to ALC, and is the number of reference points used for averaging over the design space.

Usually, $N_{ref} \propto N_{cand} \propto N$, $N_G \propto N$, and $\omega=3$. The expressions in Table \ref{tab:complexity} can thus be written as $\bigO\left(N_{mle}N^{4}+pN^3\right)$ for ALM; $\bigO\left(N_{mle}N^{4}+pN^5\right)$ for ALC, and $\bigO\left(N_{mle}N^{4}+pN^4\right)$ for MICE. Observe that ALM has a much lower computational complexity than the others, and ALC is computationally prohibitive for large $N$. MICE is computationally cheaper than ALC, as long as the ratio $(N_G-N)/N$ is not too large. In the computer experiment setting, $N_G$ can be chosen to not be too large out of computational convenience.

Because the maximum likelihood often is the most expensive step, $N_{mle}N^{1+\omega}$, a reduction in cost can be achieved by only updating the MLEs of $\sbxi$ at every $i$-th step, for some number $i$. This tends to reduce $T_{mle}$ substantially, giving MICE a significant advantage over ALC.

\begin{figure}[!ht]
\centering
\includegraphics[width=1.0\linewidth]{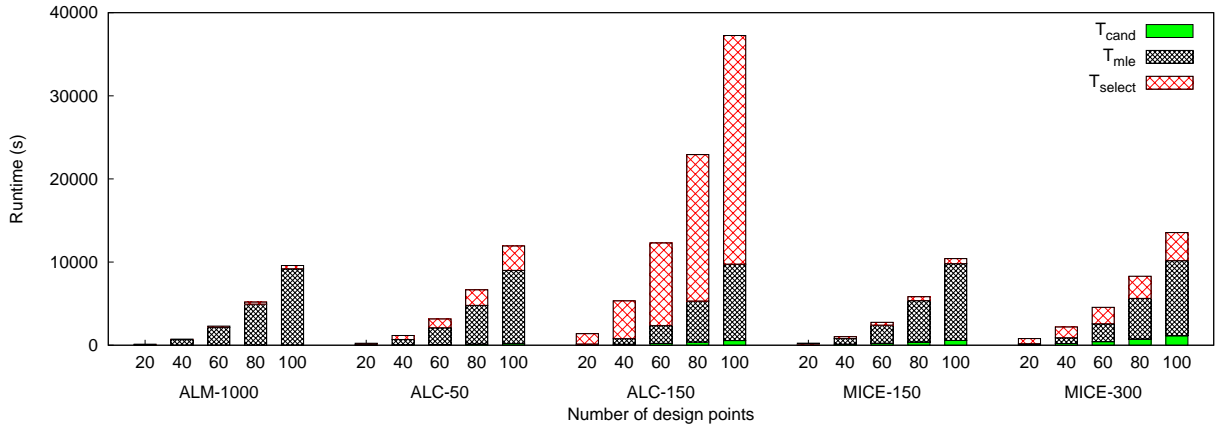}
\caption{Running time of the different sequential design algorithms for selecting designs of different fixed sizes. The study is on the Oscillatory function over $[0,1]^4$.}\label{fig:4dhist}
\end{figure}

The cost to generate candidate sets varies depending on the choice of sampling technique and the desired size. For instance, minimax designs are more computationally intensive than maximin designs \cite{BJ2011}.

\section{A numerical comparison}
\label{sec:5}
In this section, we present a numerical comparison between MICE, ALM and ALC, to better understand, as well as compare, the different sequential designs. We also consider MmLHD, which is a maximin-distance design within the class of LHDs, and mMLHD, a minimax-distance design within the class of LHDs. Note that Mm stands for maximin, and mM for minimax. MmLHDs tend to cover the parameter space better than Mm-distance designs, which are not restricted to the class of LHDs, but at the expense of lower Mm-distance scores. Hence, MmLHD can be seen as a compromise between Mm- and mM-distance designs \cite{BJ2011}. MmLHD and mMLHD select a LHD from a pool of $1000$ LHDs. The mM-distance is measured using 1000 reference points over $\sX$ on a LHD.

When training the GP emulator, all the input variables are scaled to lie in $[0,1]^p$, and all outputs are scaled to have zero mean and unit variance. The computational budget is limited to 150 design points. This budget is reasonable in realistic simulations where resolution is high. The metric of prediction accuracy is primarily the empirical RMSPE, as defined in Eq. \eqref{eq:rmspe}, against design size. The RMSPE is calculated over a 1000-point LHD. The test functions have been selected to cover different input dimension sizes and difficulty levels. The results are presented as averages of ten replicates. For each replication, a different initial design is used, consisting of two points sampled using mMLHD. All methods are compared using the same initial designs. The MmLHD and mMLHD results are averages of ten tries, and calculated for design sizes 50, 75, 100, and 120. The actual runtime is another factor that must be considered. 

A stationary GP with a Mat\'ern covariance with $\nu=5/2$ is used in all examples. Because the size of the candidate set has such a significant effect on the results, the number of candidate points are included in the method names, for example, we denoted MICE with $N_{cand}=150$ by MICE-150. With ALC the computational cost, with respect to $N_{cand}$, is substantially higher than with ALM and MICE. Hence, for ALM, we consider $N_{cand}=1000$, for MICE $N_{cand}=150,300$, and for ALC $N_{cand}=150$. ALM is kept at $N_{cand}=1000$ because its algorithm cost is low. The candidate sets are LHDs, selected based on the maximin criterion with respect to the current design.

The remaining parameters are specified as $N_{ref}=N_{cand}$ for ALC, as used in \cite{GL2009,SWGO2000}, and $\tau^2_s=1$ for MICE. We have also included results for a range of different choices of $\tau^2_s$. In particular, $\tau^2_s=10^{-12}$ which behaves as the MI algorithm, since then $\tau^2_s \approx \tau^2$.

The optimiser employed for the MLE method is a real-coded genetic algorithm \cite{DA1995} with settings that require 1024 calls to the log-likelihood. The values for the uncertain correlation parameters are fixed until the current design is of a specific size (20 if $p>$4, else 10). 

\subsection{Alan Genz's Oscillatory function} The ``Oscillatory'' function belongs to a family of test functions \cite{G1991} proposed by Alan Genz for the study of quadrature methods. The function is $y(\bx)=\cos \left( \bc \cdot \bx + 2\pi w \right), \quad \bx \in [0,1]^p$. The vector $\bc=(c_1,c_2,\ldots,c_p)$ determines the level of difficulty along the different directions of $\sX \subset \Rset^p$, and $w$ is the displacement. To study the impact of dimension size $p$ on the difficulty to predict untried points, $\bc$ is constrained as $\sum^p_{i=1} c_i=h, c_i>0$, where $h$ can be held fixed in order to maintain the difficulty level of the problem for different choices of $p$. Two case examples are considered: $\bc=(1.85,2.51,1.94,2.70)^T$ and $w=0.43$ over $[0,1]^4$, and $\bc=(0.14,1.69,0.81,1.73,2.10,0.42,0.14,1.97)$ and $w=0.4$ over $[0,1]^8$, where $h=9$.

\begin{figure}[!ht]
\minipage{0.48\textwidth}
  \includegraphics[width=1.0\linewidth]{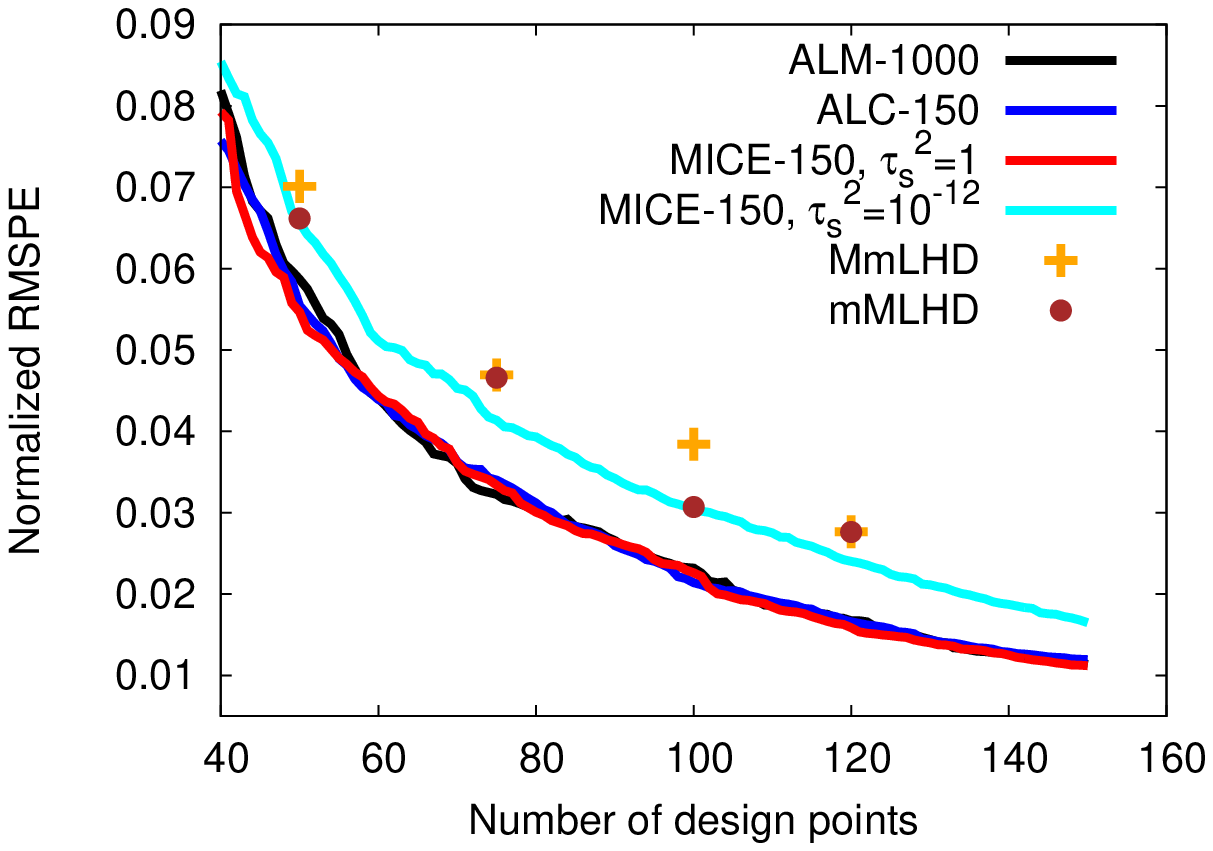}
\endminipage\hfill
\minipage{0.48\textwidth}
  \includegraphics[width=1.0\linewidth]{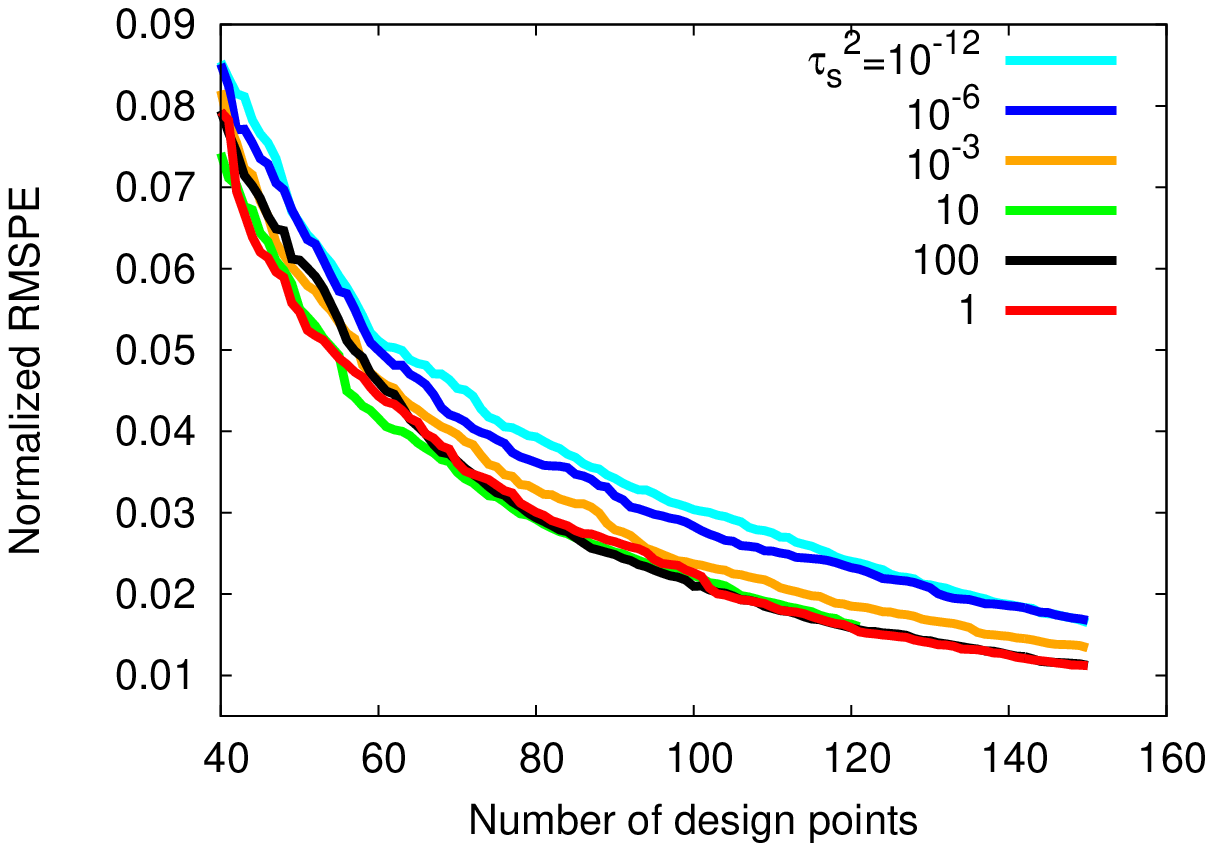}
\endminipage\hfill
\caption{Left: comparison between algorithms for the Oscillatory function over $[0,1]^4$. Right: the performance with MICE-150 for different choices of $\tau^2_s$.}\label{fig:oscillatory}
\end{figure}

\begin{figure}[!ht]
\minipage{0.48\textwidth}
  \includegraphics[width=1.0\linewidth]{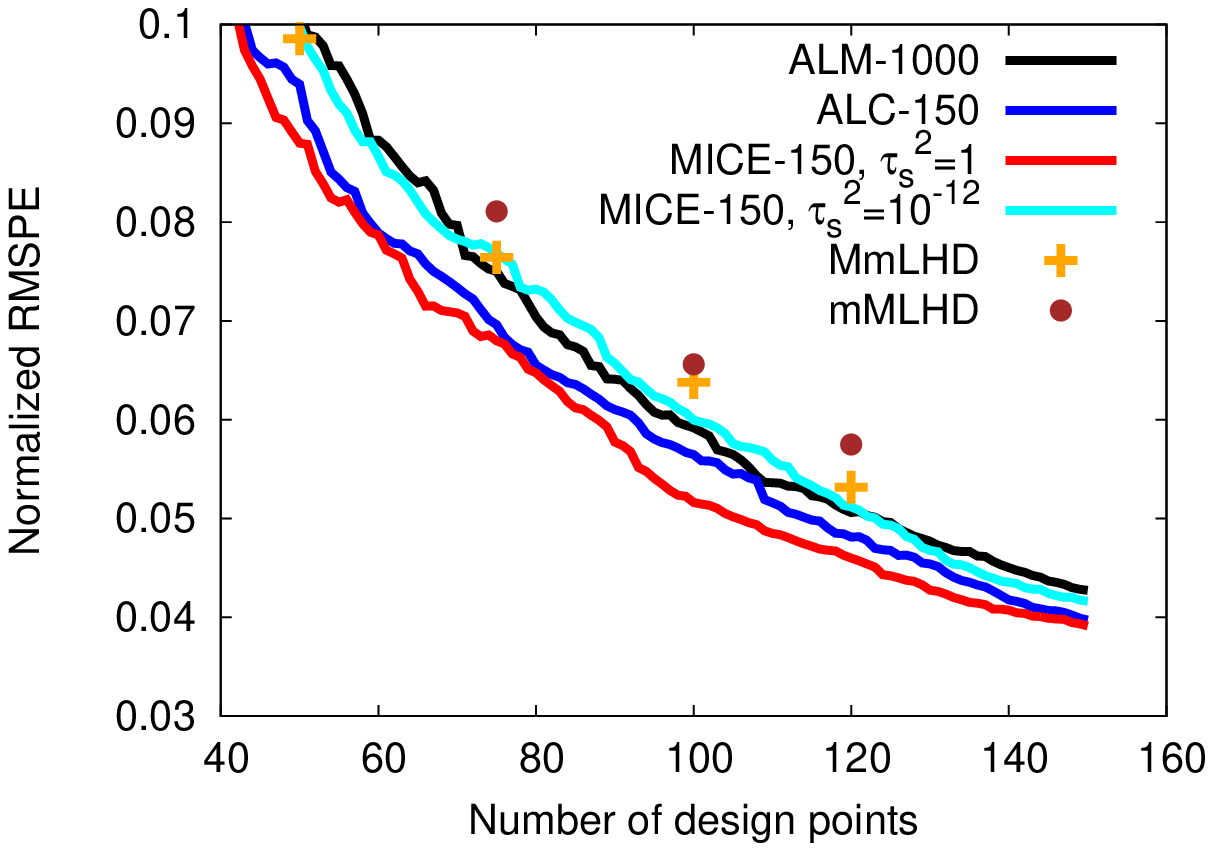}
\endminipage\hfill
\minipage{0.48\textwidth}
  \includegraphics[width=1.0\linewidth]{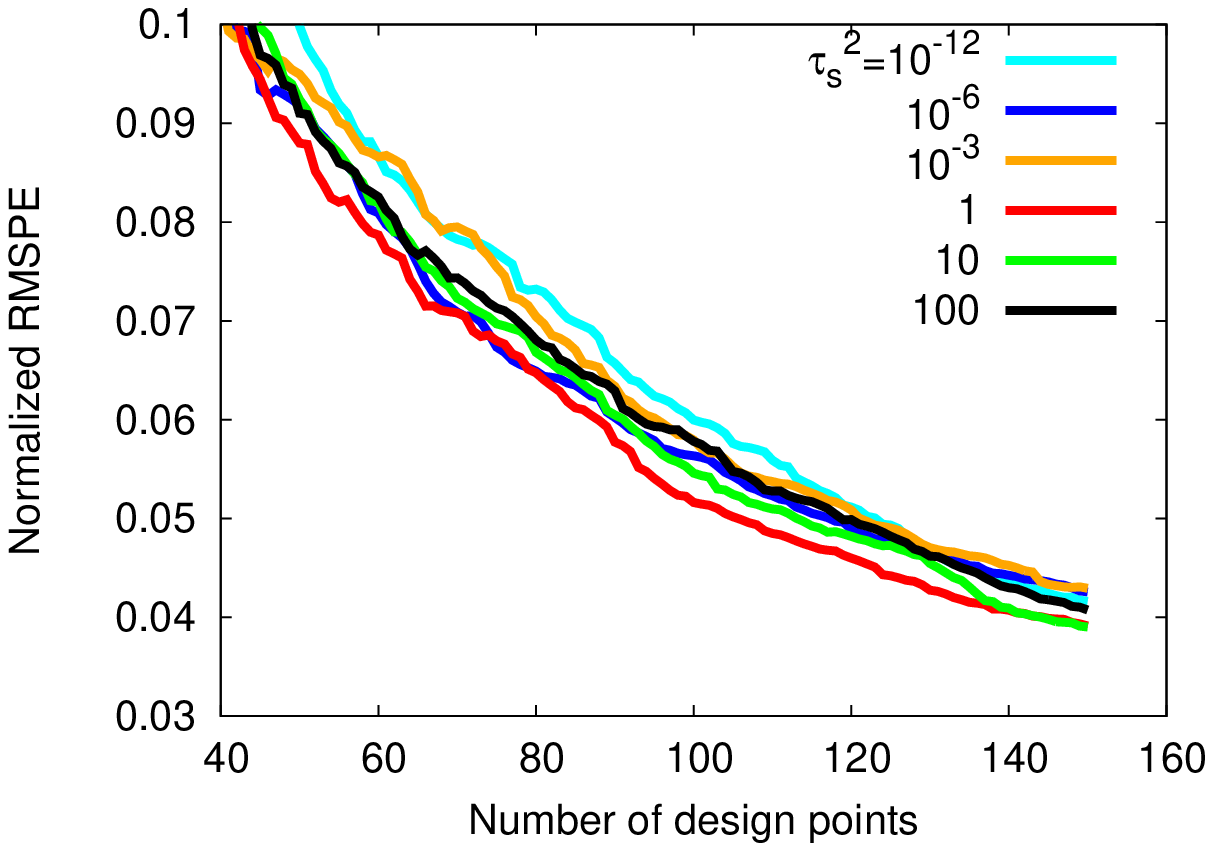}
\endminipage\hfill
\caption{Left: comparison between algorithms for the Oscillatory function over $[0,1]^8$. Right: the performance with MICE-150 for different choices of $\tau^2_s$.}\label{fig:oscillatory8D}
\end{figure}

As can be observed in Figure \ref{fig:oscillatory}, the sequential designs outperform the ones based on LHDs. As expected, since MmLHD and mMLHD, even if well spaced, do not take into account that $y(\bx)$ is anisotropic. The worst performing sequential design is MICE-150 with $\tau_s^2=10^{-12}$, which in fact uses the MI criterion. The poor performance is down to the issue discussed in Section \ref{sec:issue}. In the 4-dimensional case, ALM-1000, ALC-150, and MICE-150, produce similar results in terms of prediction error, but as shown in Figure \ref{fig:4dhist}, the time to run ALC is significantly higher than for ALM and MICE, which in many cases make it the least favorable, especially if $y(\bx)$ is cheaper to evaluate. Even if one assumes that the less costly ALC-50 would produce a similar performance as ALC-150, it would still not be competitive in this case. Observe that $\tau^2_s=1$ performs the best.

\subsection{Piston simulation function}
Here we consider a 7-dimensional example from \cite{BES2007}, where the output describes the circular motion of a piston within a cylinder; it obeys the following equations:
\begin{align*}
y(\bx) & = 2\pi\sqrt{\dfrac{x_1}{x_2+x_3^2\frac{x_4x_5}{x_6}\frac{x_7}{g_1(\bx)}}},\text{ where} &g_1(\bx) & = \frac{x_3}{2x_2}\left( \sqrt{g_2^2(\bx)+4x_2\frac{x_4x_5}{x_6}x_7}- g_2(\bx)\right) \nonumber \\
\quad & \quad &g_2(\bx) & = x_3x_4+19.62x_1-\frac{x_2x_5}{x_3} \nonumber
\end{align*}
Here $y(\bx)$ is the cycle time (s) which varies with seven input variables. The design space is given by $x_1 \in [30,60]$ (piston weight, $kg$), $x_2 \in [1000,5000]$ (spring coefficient, $N/m$), $x_3 \in [0.005,0.020]$ (piston surface area, $m^2$), $x_4 \in [90000,110000]$ (atmospheric pressure, $N/m^2$), $x_5 \in [0.002,0.010]$ (initial gas volume, $m^3$), $x_6 \in [340,360]$ (filling gas temperature, K) and $x_7 \in [290,296]$ (ambient temperature, $K$). The nonlinearity makes this deterministic computer experiment problem challenging to emulate.
\begin{figure}[!ht]
\minipage{0.48\textwidth}
  \includegraphics[width=1.0\linewidth]{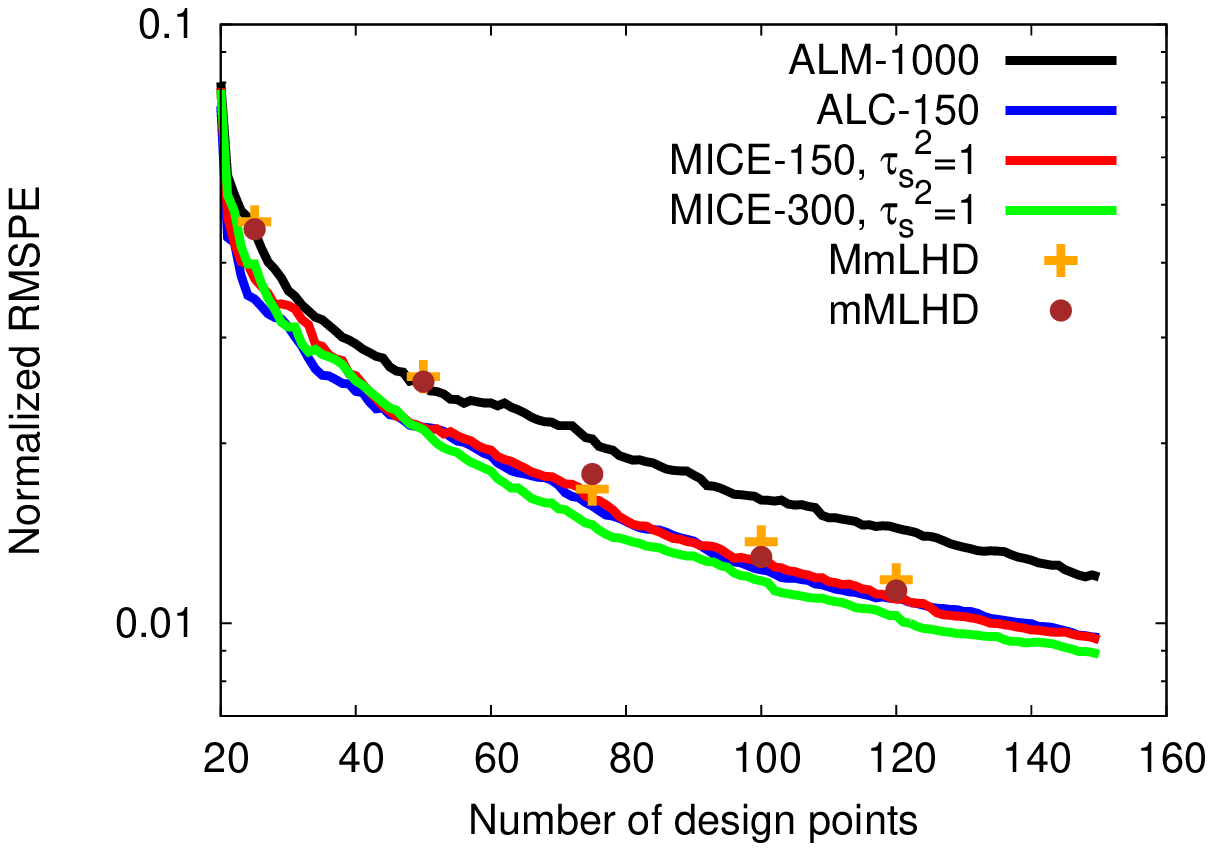}
\endminipage\hfill
\minipage{0.48\textwidth}
  \includegraphics[width=1.0\linewidth]{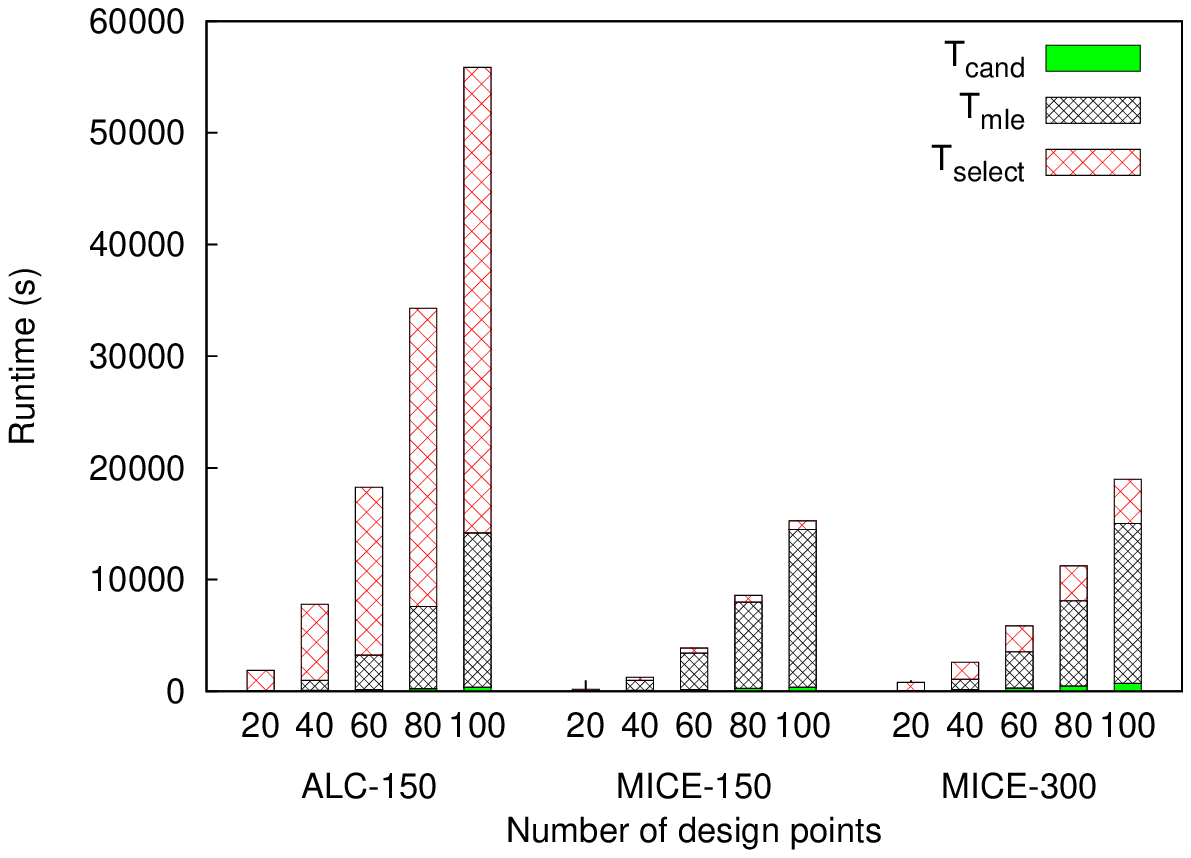}
\endminipage\hfill
\caption{Results for the 7-D Piston Simulation function.}\label{fig:piston}
\end{figure}
MICE-300 yields a slight improvement over MICE-150, see Figure \ref{fig:piston}. MICE with 300 candidate points is not that much more expensive than with 150; in fact, it is significantly cheaper computationally than ALC with 150. Again, the proposed algorithm MICE performs the best. For high-dimensional problems, ALM tends to be the worst, probably due to the high percentage of points on the boundary.

\section{Application to a tsunami simulator}
\label{sec:6}
There is a pressing need in tsunami modeling for uncertainty quantification with the specific purpose of providing accurate risk maps or issuing informative warnings. Sarri, Guillas and Dias \cite{SGD2012} were the first to demonstrate that statistical emulators can be used for this purpose. Recently, Sraj et al. \cite{SMKDH2014} studied the propagation of uncertainty in Manning's friction parameterization to the prediction of sea surface elevations, for the Tohoku 2011 tsunami event. They used a polynomial chaos (PC) expansion as the surrogate model of a low resolution tsunami simulator. Note that Bilionis and Zabaras \cite{BZ2012} showed that GP emulators can outperform PC expansions when small to moderate-sized training data are considered. Stefanakis et al. \cite{SCVDS2014} used an active experimental design approach for optimization to study if small islands can protect nearby coasts from tsunamis.

\begin{figure}[!ht]
\centering
\includegraphics[scale=0.7]{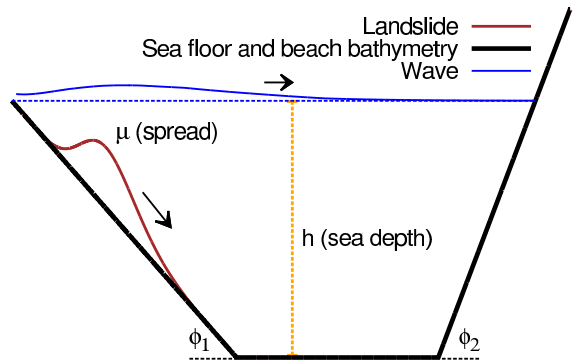}
\caption{Case example: landslide-generated tsunami event.}\label{fig:setup}
\end{figure}

We consider here the problem of predicting the maximum free-surface elevation of a tsunami wave at the shoreline, for a wide range of scenarios, following a subaerial landslide at an adjoining beach across a large body of shallow water. A tsunami wave simulator is used. A landslide of seafloor sediments, initially at the beach, has a Gaussian shaped mass distribution, and generates tsunami waves that propagates towards the opposite shoreline across from the beach (see Figure \ref{fig:setup}). The sea-floor bathymetry is changing over time, and is used as input to the tsunami simulator. The floor motion is described by the change in bathymetry of the sloping beach over time, $h(x,t)=H(x)-h_0(x,t)$, where $H(x)=x\tan{\beta}$ is the static uniformly sloping beach, and $h_0(x,t)=\delta\exp \left(-(\tilde{x}-\tilde{t})^2 \right)$ is the perturbation with respect to $H(x,t)$. Here $\tilde{x}=2\frac{x\mu^2}{\delta \tan{\phi_1}}$, $\tilde{t}=\sqrt{\frac{g}{\delta}}\mu t$, $\delta$ is the maximum vertical slide thickness, $\mu$ is the ratio of the thickness and the slide length, and $\tan{\phi_1}$ is the beach slope. The free surface elevation is defined as $z(x,t)=-h(x,t)$. It is assumed the initial water surface is undisturbed, that is, $z(x,0)=0$ for all $x$. The slope $\tan{\phi_2}$ of the beach at the opposite shoreline is chosen so that the distance between the shorelines is 2800 m. This is a shallow water problem, which means that $\tan{\phi_1}  \ll 1$, and that the translating mass movement is thin ($\mu=\delta/L \ll 1$).

We use the state-of-the-art numerical code VOLNA \cite{DPD2011} to simulate all stages of this landslide-generated tsunami event, based on nonlinear shallow water equations. We run VOLNA on a single GPU on the cluster {\em Emerald}. The bathymetry defined above is given only along one spatial coordinate, but in the code implementation of VOLNA a second spatial dimension (in this case, along the shoreline) is added to cover $10$ meters of shoreline. The mesh is defined on $[-5,5]\times[0,3000]$ (m$^2$), and consists of 312,016 triangular elements. 

We demonstrate the efficiency of the different sequential design methods for the design of a realistic computer experiments. This problem, is inspired by a benchmark problem, given at the Catalina 2004 workshop on long-wave runup models used in the validation of tsunami models. We consider 4 input parameters for emulation: $\phi_1 \in [35^\circ,70^\circ]$, $\phi_2 \in [35^\circ,70^\circ]$, $h \in [500.0,1000.
0]$, and $\mu=[0.01,0.1]$.

\begin{figure}[!ht]
\centering
\includegraphics[width=0.48\linewidth]{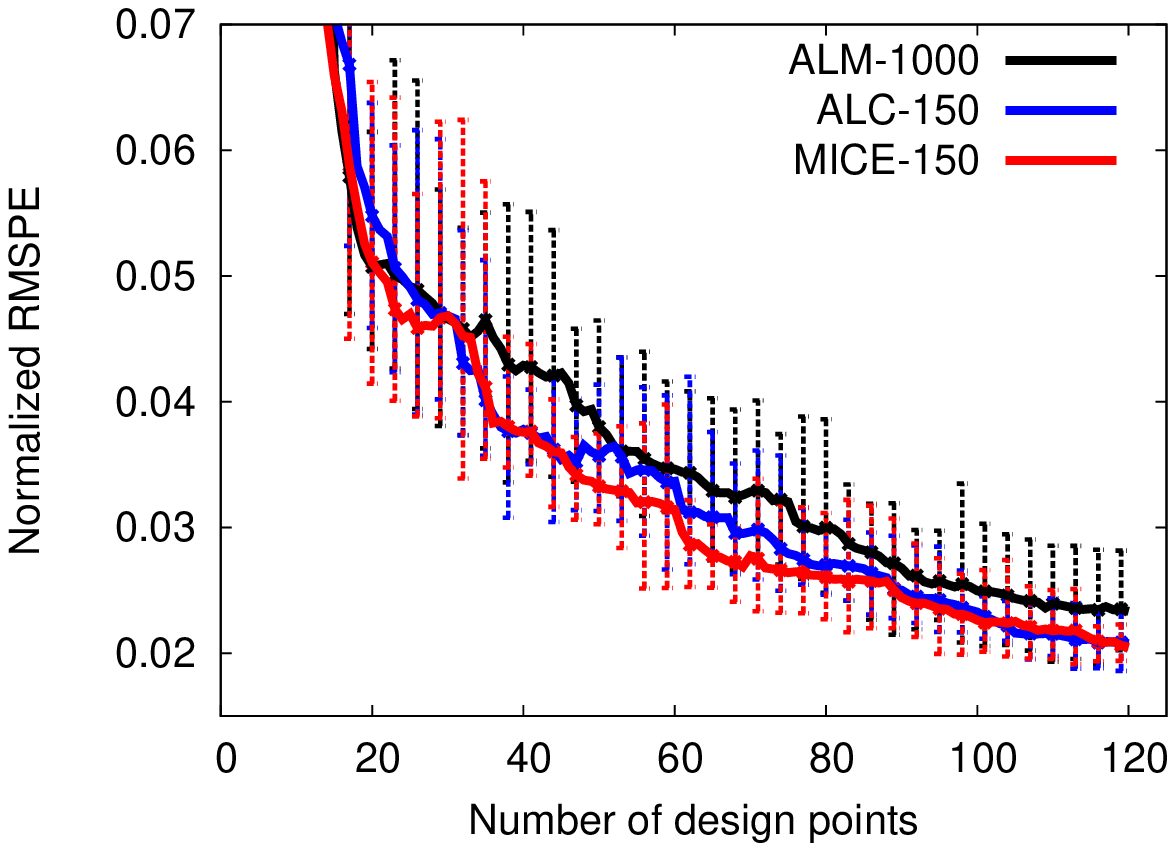}
\includegraphics[width=0.48\linewidth]{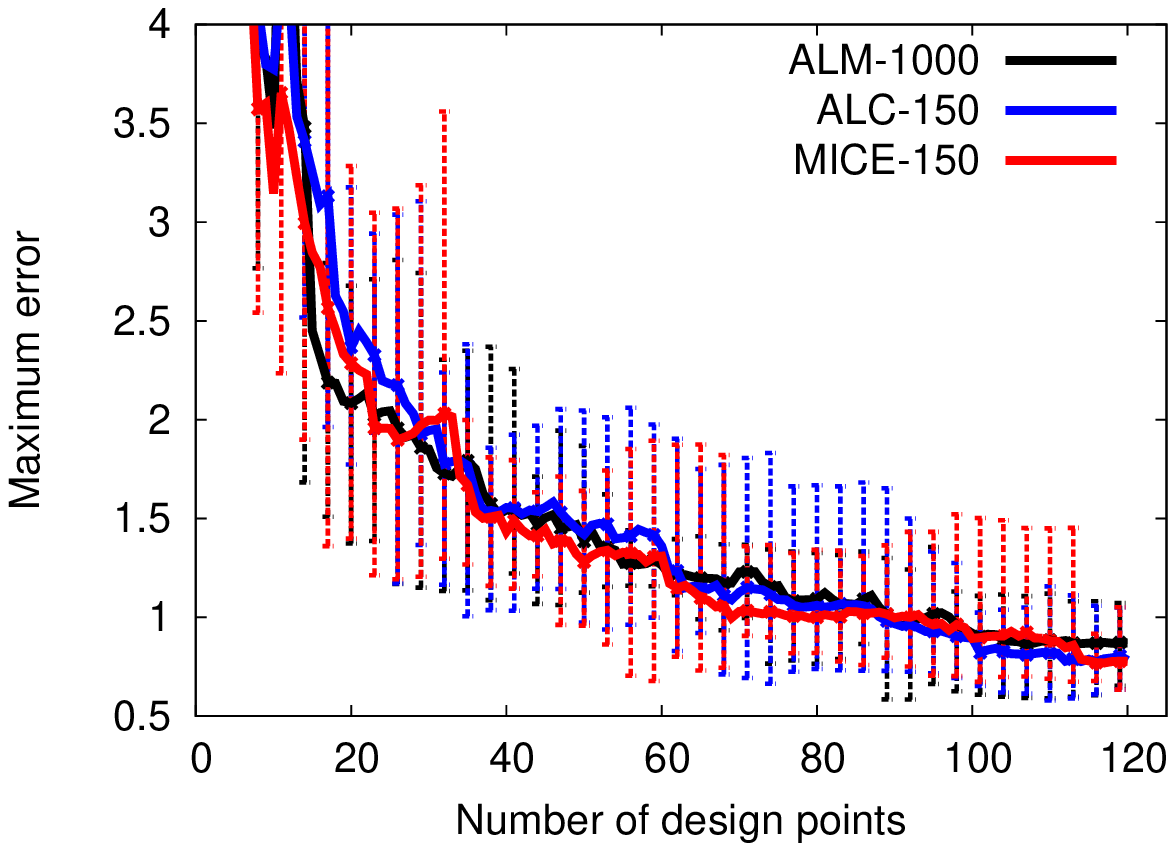}
\includegraphics[width=0.48\linewidth]{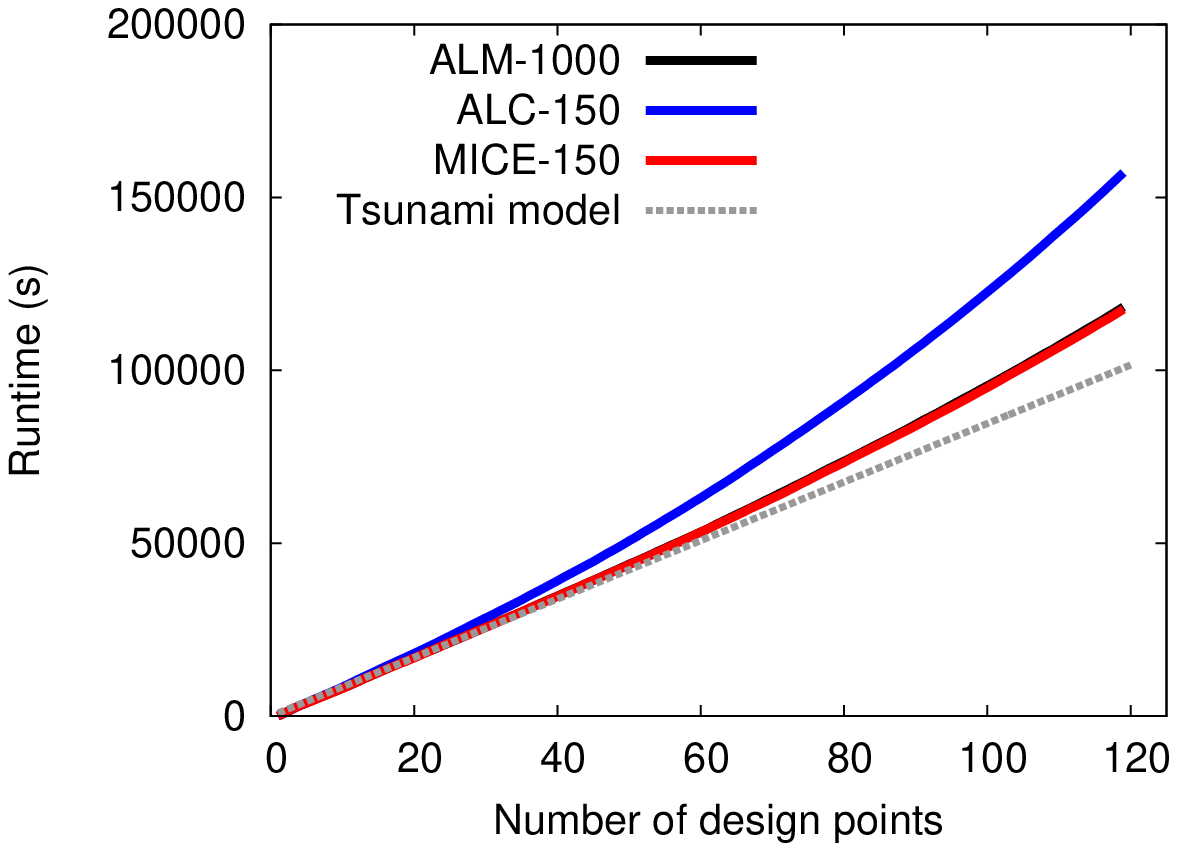}
\includegraphics[width=0.48\linewidth]{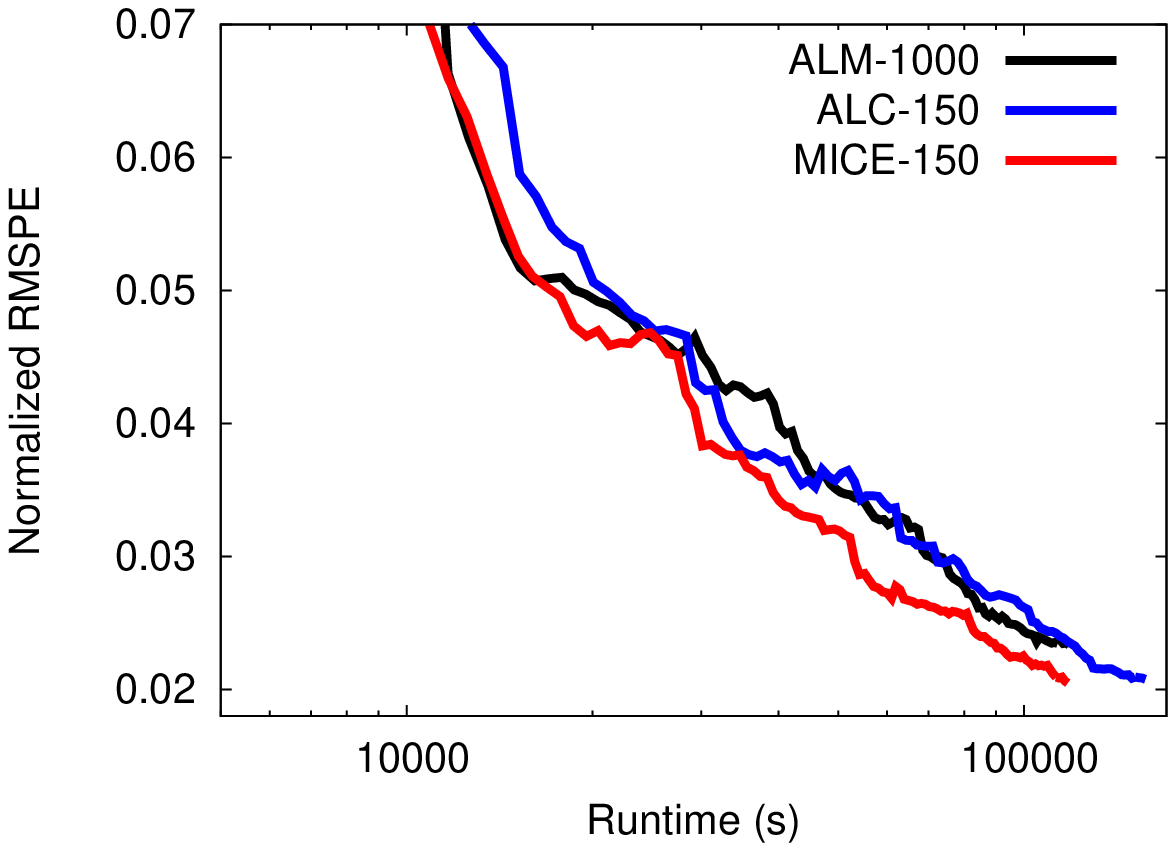}
\caption{Results for a simple tsunami model. Note the log-scale in the lower right figure.}\label{fig:tsunami4d}
\end{figure}

Some of the method-specific parameters are $N_{ref}=150$ (ALC) and $\tau^2_s=1$ (MICE). $N_{cand}=150$ is used for ALC and MI, and since ALM is relatively cheap computationally with respect to $N_{cand}$, we let $N_{cand}=1000$ for ALM. The results are averages of ten runs. As before, the GPs have a constant mean, and use the Mat\'ern covariance $\nu=\frac{5}{2}$. A hold-off set of size $500$ is used to calculate the normalized RMSPE and the maximum error. 

In Figure \ref{fig:tsunami4d} we observe that MICE performs better than ALM and ALC when considering the actual run time. ALM is more competitive when the objective is to minimize the maximum error, since it places most points on the boundary where the largest prediction errors often are located. The maximum prediction errors, over the designs of size $120$, is 1 meter, or less, in sea surface elevation for waves up to almost 10 meters. Note that in the bottom right figure the total run time is given in logarithmic scale with base 10, and the computational savings are $\sim$10-20\% by using MICE, or greater if the MLE method were to be applied more sparsely as it dominates the MICE cost. A single run of VOLNA takes on average 850 seconds. The time consumed by the simulator is represented by a gray dashed line in Figure $\ref{fig:tsunami4d}$ (bottom left figure). 

For a more realistic tsunami scenario, with more parameters, the convergence towards a good fit of the GP will be slower. Hence ALC will become relatively much more costly than MICE as ALC's cost increases steeply with the number of runs. Since each additional run will help gain a lot of precision, we expect that MICE will outperform ALC and ALM even more in such scenarios.

\section{Conclusion} 
\label{sec:7} In this paper we introduced a new mutual information-based design criterion, MICE, to find a design for good overall prediction in computer experiments. The MICE algorithm is particularly attractive in terms of time complexity of the entire design process. Our numerical studies show that, for a good range of test functions, and a realistic tsunami simulator, MICE is able to outperform popular methods such as ALC, ALM, and LHD. In addition, MICE may outperform the other designs even more (we conjecture around 50-70\% more after examining our computational summaries above, depending on the other relative costs) with less frequent updates of the MLE (e.g. every 5-10 steps) of the correlation parameters; this is something to investigate in the future in practical implementations. Our theoretical results also improve our understanding of the nugget parameter on the variance estimation, which is a key ingredient in MICE.

In this article we investigate the computational costs of the algorithms considered. The computational costs of the sequential design algorithms matter when the simulator is neither very cheap (no need for sequential design) nor extremely expensive (the cost of any algorithm is then negligible). This is generally the case in uncertainty quantification studies as models are run at a high fidelity level, but not at their highest level in order to allow the exploration of the input's influences on the outputs. If the cost of the sequential design algorithm is of the same order of magnitude as the simulator (or say 10-100 times less), then gains can be readily made by running more times the simulator, and more so when the cost of the algorithms increase steeply with design size. Furthermore, it is typical for a research project to be awarded a certain number of hours on a cluster, and thus computational complexity will increase accuracy under the same budget conditions. Another recurrent issue is that clusters are shared among many research projects, often at the local or national level. The queuing time becomes an issue as sometimes there is no cluster configuration that can accommodate the run at the time of job submission. Note that for well parallelized simulators (e.g. climate, fluid dynamics and tsunami models), the queuing time on a busy cluster can be in the order of hours, or days in some instances. By having a performant sequential design strategy, the queuing time can be reduced - sometimes dramatically in case of sudden bottlenecks - by running the simulator less times for the same accuracy.

Finally, further extensions of MICE would be welcome. One such extension would be a MICE algorithm in a non stationary setting, for example, in the treed GP form \cite{GL2009} in which subdomains of the input space, where the input-output relationships are different, are identified, and the sampling is carried out accounting for this behavior. Another possible extension would be to account for multiple outputs in terms of spatial location or behavior. Also, the desire to screen active variables along the sequential design would constitute another extension for models whose large number of variables need to be reduced before, for instance, carrying out uncertainty quantification tasks.

\section*{Appendix. Proofs of theorems}

\vspace{4px}

{\em Proof.} [Theorem \ref{theo:nugget}] 
Given a GP emulator on $D_k=(\bX_k,\by_k)$ with constant mean and a fixed correlation matrix with a nugget parameter $\tau^2$, the predictive variance for any point $\bx_i \in \bX_{k}$ can be written as:
\begin{align*}
\hat{s}^2_{\tau^2}(\bx_i)  & = \sigma^2(1-\bk^T(\bx_i))\left(\bK+\tau^2\Id\right)^{-1}\bk(\bx_i)) \\
\quad & +(\textbf{1}^T(\bK+\tau^2\Id)^{-1}\bk(\bx_i))-1)^2/(\textbf{1}^T(\bK+\tau^2\Id)^{-1}\textbf{1})),
\end{align*}
where $\Id$ is the $k \times k$ identity matrix, then
\begin{align*}
 \bk^T(\bx_i)\left(\bK+\tau^2\Id\right)^{-1}\bk(\bx_i) & = \bk^T(\bx_i)\left(\bK+\tau^2\Id\right)^{-1}(\bk(\bx_i)+\tau^2\textbf{e}_i)-\tau^2\bk^T(\bx_i)\left(\bK+\tau^2\Id\right)^{-1}\textbf{e}_i \\
 & = \bk^T(\bx_i)\textbf{e}_i-\tau^2\bk^T(\bx_i)(\textbf{e}_i^T(\bK+\tau^2\Id)^{-1})^T \\
 & = 1-\tau^2\bk^T(\bx_i)(\textbf{e}_i^T(\bK+\tau^2\Id)^{-1})^T \\
 & = 1-\tau^2\textbf{e}_i^T(\bK+\tau^2\Id)^{-1}\bk(\bx_i) \\
 & = 1-\tau^2\textbf{e}_i^T(\bK+\tau^2\Id)^{-1}\left(\bk(\bx_i)+\tau^2\textbf{e}_i-\tau^2\textbf{e}_i\right) \\
 & = 1-\tau^2+\tau^4\textbf{e}_i^T(\bK+\tau^2\Id)^{-1}\textbf{e}_i,
\end{align*}
where $\textbf{e}_i$ is the $i$-th unit vector. Similarly, $\textbf{1}^T(\bK+\tau^2\Id)^{-1}\bk(\bx_i) = 1-\tau^2\textbf{e}_i^T(\bK+\tau^2\Id)^{-1}\textbf{1}$. Insert these results into $\hat{s}^2_{\tau^2}(\bx_i)$, where $\bx_i \in \bX_k$, and we obtain
$$
\hat{s}^2_{\tau^2}(\bx_i) = \sigma^2\left(\tau^2-\tau^4\textbf{e}_i^T(\bK+\tau^2\Id)^{-1}\textbf{e}_i+\tau^4\dfrac{(\textbf{e}_i^T(\bK+\tau^2\Id)^{-1}\textbf{1})^2}{\textbf{1}^T(\bK+\tau^2\Id)^{-1}\textbf{1}}\right).\qquad
$$

{\em Proof.} [Theorem \ref{theo:bound}]
Suppose that the design space $\sX$ is a compact subset of $\Rset^p$, and discretized into a regular grid $\bX_{G} \subset \sX$ with spacing $\delta>0$. Assume the correlation function $K(\cdot,\cdot)$ is Lipschitz continuous, then there exists a constant $K_{L}>0$ such that $\vert \hat{s}^2(\bx_1)-\hat{s}^2(\bx_2) \vert \le K_L \Vert \bx_1-\bx_2 \Vert_2$ for all $\bx_1,\bx_2 \in \bX_{G}$, where $\Vert \cdot \Vert_2$ is the Euclidean norm. Suppose we have a Gaussian emulator with constant mean, and a non-negative nugget parameter $\tau^2$. Then, for any $\varepsilon>0$, assuming $\bX_{G}$ has grid spacing $\delta \le 2\varepsilon/(\sqrt{p}K_{L})$, $\hat{s}^2(\bx^*)$ is $\varepsilon$-close to $\hat{s}^2(\bx_n)$ for any untried point $\bx^* \in \sX$, where $\bx_n$ is the member of $\bX_{G}$ closest to $\bx^*$. According to Theorem \ref{theo:nugget} for any point $\bx_i \in \bX_{G}$ the predictive variance can be written as:
$$
\hat{s}^2_{\tau^2}(\bx_i) = \sigma^2\left(\tau^2-\tau^4\textbf{e}_i^T(\bK+\tau^2\Id)^{-1}\textbf{e}_i+\tau^4\dfrac{(\textbf{e}_i^T(\bK+\tau^2\Id)^{-1}\textbf{1})^2}{\textbf{1}^T(\bK+\tau^2\Id)^{-1}\textbf{1}}\right),
$$
where $\Id$ is the identity matrix, and $\textbf{e}_i$ the $i$-th unit vector. Hence, for any $\varepsilon>0$ there exists a grid spacing $\delta>0$ so that $-\sigma^2\tau^4b_1(\tau^2)-\varepsilon<\hat{s}^2_{\tau^2}(\bx^*)-\sigma^2\tau^2<\sigma^2\tau^4b_2(\tau^2)+\varepsilon$, where $b_1(\tau^2)=\max\left\{ \textbf{e}_i^T(\bK+\tau^2\Id)^{-1}\textbf{e}_i : \bx_i \in \bX_{G} \right\}$, and \\ $b_2(\tau^2)=\max\left\{ \dfrac{(\textbf{e}_i^T(\bK+\tau^2\Id)^{-1}\textbf{1})^2}{\textbf{1}^T(\bK+\tau^2\Id)^{-1}\textbf{1}} : \bx_i \in \bX_{G} \right\}$. \quad

{\em Proof.} [Theorem \ref{theo:opt}] This proof follows closely the proof of Lemma 5 and Theorem 7 in \cite{KSG2008}. Let us suppose that $\bX_{G_1} \subset \sX$ and $\bX_{G_2} \subset \sX$ are equidistant grids with spacing $2\delta$, for some $\delta>0$, and that  and that $\bX_{G_2}$ is obtained by translating $\bX_{G_1}$ by distance $\delta$ in Euclidean norm. $\bX_{G_1}$,$\bX_{G_2}$ are assumed to cover $\sX$ in terms of compactness. In the context of experimental design, let us consider $\bX_{G_1}$ to be the set of points available for selection. For a design point $\bx$ in $\bX_{G_1}$, we denote by $\tilde{\bx}$ the corresponding point in $\bX_{G_2}$, that is, $\Vert \bx-\tilde{\bx} \Vert \ge \delta, \forall \bx \in \bX_{G_1}$. Let us denote by $\bar{Y}_1,\bar{Y}_2$ the restriction of the GPs to $\bX_{G_1},\bX_{G_2}$, respectively, and, for a random variable $\cdot$ in $\bar{Y}_1$, we denote by $\tilde{\cdot}$ the corresponding translated random variable in $\bar{Y}_2$. Also, $\sX$ is compact and $K(\cdot,\cdot)$ is continuous; hence $\vert K(\bx,\bx')-K(\tilde{\bx},\tilde{\bx}') \vert \le \varepsilon_1, \forall \bx,\bx' \in \bX_{G_1}$ ($K(\cdot,\cdot)$ uniformly continuous over $\sX$). Let $\bX_k$ be a subset of $\bX_{G_1}$. For any $\bx \in \bX_{G_1} \backslash \bX_k$, we assume that $\Hop(\bx \vert \bX_k) \ge \Hop(\bx \vert \tilde{\bX_k})$ for $\vert \bX_k \vert \le 2N$, which is empirically justified in \cite{KSG2008}. 

Let $\bX_k$ be a subset of $\bX_{G_1}$, and consider a GP on $D_k=(\bX_k,\by_k)$ with a nugget parameter $\tau^2_1>0$, and a GP emulator on $\bX_{G_1} \backslash \bX_{k} \subseteq \bX_{G_2}$ with a nugget $\tau^2_2>0$. First, let us determine an upper bound for $\vert \hat{s}^2_{k}(\bx)-\hat{s}^2_{G_1 \backslash k}(\bx) \vert$: 
\begin{align*}
 & \vert \hat{s}^2_{k}(\bx)-\hat{s}^2_{G_1 \backslash k}(\bx) \vert = \sigma^2 \vert \bk^T_k(\bx)\bK_k^{-1}\bk_k(\bx)-\bk^T_{G_1 \backslash k}(\bx)\bK^{-1}_{G_1 \backslash k}\bk_{G_1 \backslash}(\bx) \vert \\
 & \le \sigma^2 ( \Vert \bk^T_k(\bx)-\bk^T_{G_1 \backslash k}(\bx) \Vert_2\Vert \bK^{-1}_k \Vert_2(\Vert\bk_k(\bx)\Vert_2+\Vert\bk_{G_1 \backslash k}(\bx)\Vert_2)  \\
 & + \Vert \bk^T_k(\bx) \Vert_2 \Vert \bK^{-1}_{k}-\bK^{-1}_{G_1 \backslash k} \Vert_2 \Vert \bk^T_{G_1 \backslash k}(\bx) \Vert_2)
\end{align*}
Since $K(\cdot,\cdot)$ is uniformly continuous over $\sX$, we know that $\forall \varepsilon_1>0$ there exists a spacing $\delta>0$ such that, for $\Vert \bx - \tilde{\bx} \Vert \le \delta$, $\vert K(\bx,\bx')-K(\bx,\bx') \vert \le \varepsilon_1$ for $\bx \neq \bx'$, and $\Vert \bK_{k,\tau^2_1}-\bK_{G_1 \backslash k,\tau^2_2} \Vert_2 \le \sqrt{N}N\varepsilon_1+\sqrt{N}\vert \tau^2_1-\tau^2_2 \vert$. We also derive $\Vert \bk^T_k(\bx)-\bk^T_{G_1 \backslash k}(\bx) \Vert_2 \le \varepsilon_1\sqrt{N}$, and similarly, $\Vert \bk^T(\bx) \Vert_2 \le C\sqrt{N}$, where $\Vert \cdot \Vert_2$ is the Euclidean norm, and $C=\max_{\bx \in \sX} K(\bx,\bx)$. We assume wlog that $C=1$. Furthermore:
\begin{align*}
\Vert \bK^{-1}_{k,\tau_1^2}-\bK^{-1}_{G_1 \backslash k,\tau_2^2} \Vert_2 & = \Vert \bK^{-1}_{k,\tau^2_1}(\bK_{k,\tau^2_1}-\bK_{G_1 \backslash k,\tau^2_2})\bK^{-1}_{G_1 \backslash k,\tau^2_2} \Vert_2  \\
& \le \Vert \bK^{-1}_{k,\tau^2_1} \Vert_2\Vert \bK_{k,\tau^2_1}-\bK_{G_1 \backslash k,\tau^2_2} \Vert_2 \Vert \bK^{-1}_{G_1 \backslash k,\tau^2_2} \Vert_2 \\
\qquad & \le (1+\tau_1^{2})^{-1}(1+\tau_2^{2})^{-1}\sqrt{N}\left(N\varepsilon_1+\vert \tau_1^2-\tau_2^2 \vert\right) \le \sqrt{N}N\varepsilon_1+\sqrt{N}\vert \tau_1^2-\tau_2^2 \vert,
\end{align*}
where we used that $\bK$ is positive semidefinite, which means that $\Vert \bK^{-1} \Vert_2 = \lambda_{min}(\bK)^{-1} \le (1+\tau^2)^{-1}$, where $\lambda_{min}(\bK)$ is the smallest eigenvalue. We thus obtain the following bound:
\begin{align*}
\vert \hat{s}^2_{k,\tau^2_1}(\bx)-\hat{s}^2_{G_1 \backslash k,\tau^2_2}(\bx) \vert & \le \sigma^2(2\varepsilon_1N(1+\tau_1^{2})^{-1}+N(1+\tau_1^{2})^{-1}(1+\tau_2^{2})^{-1}\sqrt{N}(N\varepsilon_1+\vert \tau_1^2-\tau_2^2\vert)) \\
& \le \sigma^2(2\varepsilon_1N+N\sqrt{N}(N\varepsilon_1+\vert \tau^2_1-\tau^2_2 \vert)).
\end{align*}
Then, for any $\varepsilon>0$ we can choose the grid spacing $\delta>0$ such that $\varepsilon \ge \varepsilon_1\tau_2^{2}\sigma^2N(2N+N^{3/2})$. Hence, $\vert \hat{s}^2_{k,\tau_1^2}(\bx)-\hat{s}^2_{G_1 \backslash k,\tau_2^2}(\bx) \vert \le \varepsilon\tau^2_2+\sigma^2N^{3/2}\vert \tau^2_1-\tau^2_2 \vert$, and, in turn, 
\begin{align*}
& \Hop_{\tau^2_1}(\bx \vert \bX_k)-\Hop_{\tau^2_2}(\bx \vert \bX_{G_1} \backslash \bX_{k})=\frac{1}{2}\log\left(\frac{\hat{s}^2_{k,\tau^2_1}(\bx)}{\hat{s}^2_{k,\tau^2_2}(\bx)}\right) \\
& =\frac{1}{2}\log\left(1+(\hat{s}^2_{k,\tau^2_1}(\bx)-\hat{s}^2_{G \backslash k,\tau^2_2}(\bx))/\hat{s}^2_{G_1 \backslash k,\tau^2_2}(\bx)\right) \\
& \le \frac{1}{2}\log\left(1+\varepsilon+N^{5/2}\vert \tau_1^2-\tau_2^2\vert/\tau^2_2\right) \le \varepsilon+N^{5/2}\vert \tau_2^2-\tau_1^2\vert/\tau^2_2.
\end{align*}
We used that $\hat{s}^2_{G_1 \backslash k}(\bx)\ge\sigma^2\tau_2^2/N$ (see Theorem \ref{theo:nugget}). Suppose that estimates are available for the correlation parameters $\sbxi$; replacing $K(\bx,\bx)$ by $K(\bx,\bx';\hat{\sbxi})$ throughout the calculations above. Then, an extra term is added to $\hat{s}^2(\bx)$ to account for the parameter uncertainty \cite{ZC1992}: $\hat{s}^2(\bx;\hat{\sbxi})=\sigma^2(1-\bk^T(\bx;\hat{\sbxi}_i)\bK_{\hat{\sbxi}_i}^{-1}\bk(\bx;\hat{\sbxi}_i))+E\left( (\hat{y}(\bx;\sbxi)-\hat{y}(\bx;\hat{\sbxi}_i))^2 \right)$. The estimates are updated at each greedy step, denoted by $\hat{\sbxi}_i$, for greedy step $i$. Using Eq. \eqref{eq:predmean}, with zero-mean, $\hat{y}(\bx;\sbxi)-\hat{y}(\bx;\hat{\sbxi}_i) = \bk^T_{\sbxi}(\bx)\bK_{\sbxi}^{-1}\by_k-\bk^T_{\hat{\sbxi}_i}(\bx)\bK_{\hat{\sbxi}_i}^{-1}\by_k$. Let us assume that $\Vert \by_k \Vert_2 \le \sqrt{N}$ (normalized). We know that there exists a constant $\alpha\ge0$ such that, for all $\{\hat{\sbxi}_i\}^k_{i=1}$, and for all, $\bx,\bx' \in \sX, \vert K(\bx,\bx';\sbxi)-K(\bx,\bx';\hat{\sbxi}_i) \vert \le \alpha$. Then, $E(\hat{y}(\bx;\hat{\sbxi}_i)-(\hat{y}(\bx;\sbxi))^2) = E( (\bk^T_{\hat{\sbxi}_i}(\bx)\bK_{\hat{\sbxi}_i}^{-1}\by_k-\bk^T_{\sbxi}(\bx)\bK_{\sbxi}^{-1}\by_k)^2) \le E((\Vert\bk^T_{\sbxi}(\bx)-\bk^T_{\hat{\sbxi}_i}(\bx)\Vert_2\Vert \bK_{\hat{\sbxi}_i}^{-1} \Vert_2\Vert \by_k \Vert_2+\Vert \bk_{\hat{\sbxi}_i}^T(\bx)\Vert_2\Vert\bK^{-1}_{\sbxi}-\bK^{-1}_{\hat{\sbxi}_i}\Vert_2)\Vert\by_k\Vert_2)^2) \le \alpha^2N^2(1+N^{3/2})^2$. As a result, using similar calculations, $\Hop(\bx \vert k)-\Hop(\bx \vert k,\hat{\sbxi})\le\frac{1}{2}\log((\hat{s}_k^2(\bx)+\alpha^2N^2(1+N^{3/2})^2)/\hat{s}_D^2(\bx))\le\frac{1}{2}\log(1+(\alpha\sigma^{-1}\tau^{-1})^2N^3(1+N^{3/2})^2)$. Hence,
\begin{align*}
& \Hop(\bx \vert k,\hat{\xi},\tau_1^2)-\Hop(\bx \vert (\bX_{G_1} \backslash \bX_{k}),\hat{\xi},\tau_2^2)=(\Hop(\bx \vert k,\tau_1^2)-\Hop(\bx \vert (\bX_{G_1} \backslash \bX_{k}),\tau_1^2)) \\
& +(\Hop(\bx \vert (\bX_{G_1} \backslash \bX_{k}),\tau_1^2)-\Hop(\bx \vert (\bX_{G_1} \backslash \bX_{k}),\hat{\xi},\tau_1^2)) +(\Hop(\bx\vert k,\hat{\xi},\tau_1^2)-\Hop(\bx\vert k,\tau_1^2)) \\ & +(\Hop(\bx\vert (\bX_{G_1} \backslash \bX_{k}),\hat{\xi},\tau_1^2)-\Hop(\bx\vert (\bX_{G_1} \backslash \bX_{k}),\hat{\xi},\tau_2^2)) \\
& \le \varepsilon+2(\alpha\sigma^{-1}\tau^{-1})^2N^3(1+N^{3/2})^2+N^{5/2}\vert \tau_2^2-\tau_1^2\vert/\tau^2_2.
\end{align*}
The two GPs on $\bX_k$ and $\bX_{G_1} \backslash \bX_{k}$, respectively, use the same estimates $\hat{\sbxi}$. Finally, by following the same the proof of Theorem 7 in \cite{KSG2008}, we can easily get the result of this theorem.\qquad

\section*{Acknowledgments}
This research was part funded by the NERC research programme “Probability, Uncertainty and Risk in the Environment” (PURE), grant NE/J017434/1. The authors would like to acknowledge that the work presented here made use of the Emerald High Performance Computing facility made available by the Centre for Innovation. The Centre is formed by the universities of Oxford, Southampton, Bristol, and University College London in partnership with the STFC Rutherford-Appleton Laboratory.

\bibliographystyle{plain}
\bibliography{beckguillas2014}

\end{document}